# Hamiltonian variational formulation of three-dimensional, rotational free-surface flows, with a moving seabed, in the Eulerian description

by

C.P. Mavroeidis([1]) and G.A. Athanassoulis([1])(*)
con.mavroeidis@gmail.com makathan@gmail.com

([1]) School of Naval Architecture and Marine Engineering,
National Technical University of Athens, Greece

(*) Correspondence: makathan@gmail.com

**Table of contents**

**Hamiltonian variational formulation of three-dimensional, rotational free-surface flows, with a moving seabed, in the Eulerian description**

## Abstract

Hamiltonian variational principles provided, since 60s, the means of developing very successful wave theories for nonlinear free-surface flows, under the assumption of irrotationality. This success, in conjunction with the recognition that almost all flows in the sea are not irrotational, raises the question of extending Hamilton's Principle to rotational free-surface flows. The equations governing the fluid motion within the fluid domain, in the Eulerian description, have been derived by means of Hamilton's Principle since late 50s. Nevertheless, a complete variational formulation of the problem, including the derivation of boundary conditions, seems to be lacking up to now. Such a formulation is given in the present work. The differential equations governing the fluid motion are derived as usually, starting from the typical Lagrangian, constrained with the conservation of mass and the conservation of fluid parcels' identity. To obtain the boundary conditions, generic differential-variational constraints are introduced in the boundary variational equation, leading to a reformulation which permits us to derive both kinematic and dynamic conditions on all boundaries of the fluid, including the free surface. An interesting feature, appearing in the present variational derivation of kinematic boundary conditions, is a dual possibility of obtaining either the usual kinematic condition (the same as in irrotational flow) or a condition of different type, corresponding to zero tangential velocity on the boundary. The deeper meaning and the significance of these findings seem to deserve further analysis.

## 1. Introduction

### - *Motivation*

The significance of understanding and predicting phenomena related with the wave motion in the sea can be hardly overestimated nowadays. It suffices, perhaps, to mention that they are of central interest in Naval, Coastal, and Offshore Engineering, and also of great importance for oceanography. A distinctive feature of these nonlinear waves is the presence of the unknown free surface, which has to be predicted simultaneously with the underlying wave field. This feature greatly complicates both the mathematical analysis and the numerical computation of free-surface problems.

It turns out that these difficulties become milder and easier-to-handle under specific mathematical formulations. For example, in the case of an ideal (inviscid) incompressible liquid undergoing an irrotational wave motion, the Hamiltonian variational formulation, initiated by Petrov (1964) [1] and Zakharov (1968) [2], has produced effective equations for studying non-linear waves and very efficient schemes for numerical computations. The closely related unconstrained variational principle of Luke (1967) [3] facilitated further development of nonlinear wave theories and numerical techniques. In the last 60 years, there is a huge amount of works developing and exploiting irrotational wave theories, those exhibiting a Hamiltonian structure being the most successful.

However successful the irrotational model may be, it is well known that rotationality is always present in the sea. And this is true, not only for phenomena like wave breaking and air-sea interaction, but also for mild realistic sea waves. Oceanographers introduced the concept of the wave-induced non-breaking turbulence in 2004 [4, 5] and found significant improvements in upper-ocean circulation predictions when they incorporated it in general circulation models [6]. The authors of the latter paper, trying to explain the scarcity of rotational wave theories, argue that "*the success of potential theories of nonlinear waves … made the applications of non-potential wave theories seem redundant and eventually even led to them being nearly forgotten*."

### - *History and background literature*

The need for rotational wave theories, in conjunction with the extraordinary success of the Hamiltonian variational formulation for irrotational waves, raises the question of extending the method to the more realistic (and more complicated) case of rotational free-surface flows. In this case, however, the choice between Lagrangian and Eulerian description of the fluid motion becomes significant. In the second approach, the physical fields, say the velocity $\boldsymbol{u}$, the density $\rho$ and the pressure $p$, are expressed as functions of spatial coordinates $(\boldsymbol{x}, z) = (x_1, x_2, z)$ and time $t$, while in the former one the main field is the position of fluid parcels, $\boldsymbol{X}(\boldsymbol{a}, t)$, considered as a function of their initial positions $\boldsymbol{a} = (a_1, a_2, a_3)$ and the time. It is thus clear that the Lagrangian description, being non convenient in applications and almost forgotten in engineering hydrodynamics, is directly amenable to methods of Analytical Mechanics, based on virtual displacements. Lagrange himself derived the hydrodynamic equations (in the Lagrangian description) by using D'Alembert Principle of virtual work in 1815 [7] (Sec. 11), but their derivation from Hamilton's Principle should be waiting for more than a century before appearing in 1929 [8]. In the present paper, we are primarily interested in the variational formulation of rotational flows, in their Eulerian description; the Lagrangian one will be touched upon only at the extend it is necessary for developing the former.

Direct application of Hamilton's Principle to the derivation of equations governing rotational flows in the Eulerian description, stumbles upon a fundamental controversy. The validity of Hamilton's Principle is crucially dependent on applying virtual variations of the positions of fluid parcels, $\boldsymbol{X}(\boldsymbol{a},t)$, for fixed $(\boldsymbol{a},t)$. However, fluid parcels' positions are completely absent in the Eulerian formulation, and the physical fields, say $\boldsymbol{u}=\boldsymbol{u}(\boldsymbol{x},z,t)$, are defined in terms of the spatial variables. Accordingly, the natural variations of the involved fields, say $\delta\boldsymbol{u}(\boldsymbol{x},z,t)$, are variations for fixed $(\boldsymbol{x},z,t)$. The situation described above will be referred to in the sequel as the *variational controversy* (of Eulerian fluid dynamics).

The first attempt to derive the Eulerian equations of fluid dynamics by means of Hamilton's Principle was made by Herivel in 1955 [9]. He recognized the need of introducing constraints in the standard Lagrangian function (kinetic minus potential energy), and he implemented as constraints the conservation of mass and the conservation of entropy. The obtained variational equations apparently produce Euler's equation of momentum; however, the underlying velocity representation turns out to be restricted to irrotational flows, when the entropy is assumed to be constant, as it should ([1]); see also the discussion in [10] (p. 5). Let it be noted that, from a purely mechanical point of view, entropy should not be of significance for the dynamics of an ideal mechanical system, as the considered fluid flow is. The variational controversy has been resolved, at least for the bulk motion of the fluid, by means of the clever proposal of Lin [11, 12] ([2]) who introduced the purely mechanical constraint of the conservation of fluid parcels' identity. See also Sec. 3.1. Although Lin's constraint is commonly used in conjunction with the entropy constraint [10, 13], in fact, it can replace the latter, leaving us with a purely mechanical variational formulation.

The Herivel-Lin approach became the standard one after the publication of the works of Serrin [11] and Eckart [14]. Many papers have appeared since then, exploiting various aspects and deepening our understanding of this variational formulation. Penfield [15], Bretherton [16] and Salmon [17] discussed the importance of using Hamilton's Principle (the Herivel-Lin approach) for studying Eulerian rotational flows of ideal fluids. The necessity of introducing Lin's constraint has also been highlighted, by means of transformations between the two descriptions of fluid flow [17, 18]. In an alternative direction, the Eulerian variational formulation was obtained from that in the Lagrangian description, by using canonical transformations [19, 20]. The papers by Seliger and Whitham [10] and Fukagawa and Fujitani [13] dealt with many aspects of Hamilton's Principle for rotational flows, including its relation with the Clebsch approach (see below, at the end of the Introduction). They also discussed the variational formulation with reduced versions of Lin's constraint, that is, velocity representation with fewer potentials; see, also, the discussion in [17] (Sec. 5). However, as first illustrated in [16], and later proved with more advanced mathematical tools in [21–23], these consideration apply only to a restricted class of flows, with zero helicity and without points of vanishing vorticity; see, also [24] and [25] (Sec. 6.17).

All papers mentioned above, dealing with the Herivel-Lin Hamiltonian variational formulation of three-dimensional (3D) rotational flows, do not touch upon the issue of boundary conditions. In fact, the only work that the present authors found, with some discussion on boundary conditions in this context, is the book by Berdichevsky [26] (Sec. 9.3), where he derives a form of the free-surface dynamic condition, having all the kinematic conditions a priori imposed. In a different direction, which uses Hamilton's Principle in conjunction with the constrained variations of the Euler-Poincaré framework, a limited number of works dealing with free-surface boundary conditions appeared recently (see [27, 28], and references therein). Although these works are interesting and illuminative regarding various aspects of the variational formulation of unsteady vortical free-surface flows, they are limited to two-dimensional (2D) flows, utilizing a stream function. Again, kinematic conditions need to be imposed as constraints in the variational formulation.

---

([1]) The variational controversy was not well understood in 1955.

([2]) This proposal has first appeared in [11], with reference to an unpublished note of Lin. A similar reference is given in [14].

*- Contribution of the present paper*

The goal of the present paper is to provide a complete derivation of the equations of motion and of the boundary conditions for 3D rotational flows with a free surface and a moving seabed, by means of Hamilton's Principle. The fluid is assumed to be barotropic, which permits us to consider it as a purely mechanical system, avoiding the use of entropy and other thermodynamic concepts. To achieve this goal, we start with Herivel-Lin's version of Hamilton's Principle (the standard Lagrangian, constrained by the conservation of mass and the conservation of fluid parcels' identity), calculate the total variation of the actual functional with respect to all functional arguments (physical fields and fields of constraints), and deduce the equations of motion within the 3D fluid domain. Up to this point we follow the standard approach, as in [10], with the minor difference that we do not use entropy nor the entropic constraint. As usual, the Clebsch velocity representation, $\boldsymbol{u} = -\nabla k + \sum A_i \nabla a_i$, where $k$, $A_i$, $a_i$ are scalar functions (Clebsch potentials), is derived variationally. It is also observed that the number of pairs $A_i$, $a_i$, entering into the velocity representation, do not affect the variational treatment. A possible interpretation of this situation may be that the velocity representation belongs to the underlying kinematics and, thus, it is not fully controlled by the variational principle.

The calculation of the total variation of the action functional provides us with boundary terms as well, from which it is expected to derive the boundary conditions. This seems not to be possible, however, if we consider the boundary variations of the involved fields to be independent on the boundary, as it happens within the fluid domain. This controversy is addressed, by the present authors, to the fundamental variational controversy of the Eulerian fluid dynamics. The main argument (conjecture) is that Lin's constraint, expressed by volume integrals, does not effectively act on the boundary, which is a lower-dimensional manifold. To cope with this difficulty, use is made of differential-variational constraints, which express the local Eulerian field variations by means of the virtual displacement $\delta \boldsymbol{X}$ of the fluid parcels. The latter are pointwise valid, thus valid on the boundary as well. After substituting the variations of the Eulerian fields on the boundary with the corresponding expressions in terms of $\delta \boldsymbol{X}$, the number of independent boundary terms decreases and their new forms, in conjunction with standard variational arguments, provide us with the expected boundary conditions on the free surface, on the moving seabed, and on any lateral rigid-wall boundary. These results can be considered as an a posteriori justification of our conjecture, on the inadequacy of Lin's volume-integral constraint for the boundary conditions. An interesting and rather unexpected feature, appearing in the variational derivation of kinematic boundary conditions, is the possibility of a (non-exclusive) duality. That is, the variational equation supports two types of kinematic conditions: either the usual kinematic condition (the same as in irrotational flow) or a condition of different type, corresponding to zero tangential velocity on the boundary. The deeper meaning and the significance of these findings seem to deserve further analysis.

The paper is organized as follows: In Sec. 2, we describe the geometry of the fluid domain, and present the usual differential formulation (governing equations and boundary conditions) of the problem. The Hamiltonian action functional, along with appropriate constraints, are presented in Sec. 3. Both the usual integral constraints and additional, differential-variational constraints are discussed in this section. The general variational equation is derived in Sec. 4, and it is exploited for obtaining the equations of motion within the fluid domain, including explicit representations of the velocity and pressure fields in terms of Clebsch potentials. Sec. 5 is devoted to the treatment of the boundary variational equation. Its primitive form, as obtained by the standard approach of the Calculus of Variations, is reformulated by using the differential-variational constraints, resulting in a new form from which the complete set of boundary conditions is derived, for each part of the boundary (free surface, moving seabed, fixed lateral rigid-wall boundaries). Last, in Sec. 6 we discuss the findings of this work and arrive at some concluding remarks.

Before concluding this introductory section, we believe it is appropriate to discuss some

- *Related research directions which are not considered in this work*

In fact, the history of the variational formulation for rotational flows in the Eulerian description started in the 19th century, with the pioneering work of Clebsch in 1859 [29]. In this work, the author applies the theory of Pfaffian forms to obtain a representation of the velocity field in terms of three scalar functions (called Clebsch potentials), and observes that the hydrodynamic equations (expressed in terms of these potentials) may be obtained as the extremal condition of an integral functional. This functional is essentially the space-time integral of the pressure field. The main points of the Clebsch approach were summarized by Bateman (1929, 1944) [30, 31] in a form adapted to compressible flows. Seliger and Whitham [10] and Fukagawa and Fujitani [13] have discussed the relation of the Clebsch approach with Hamilton's Principle, without considering boundary conditions. Luke [3] provided a short note suggesting the possibility of exploiting the Clebsch-Bateman variational principle for barotropic free-surface flows, and this suggestion was recently elaborated further by Timokha [32]. In this approach, the dynamic free-surface condition is easily obtained, because the Lagrangian density is just the pressure field. We do not follow this line of thought in the present paper, choosing instead the Hamilton's Principle as a basis of our search for the variational formulation, which provides direct connection with the foundational ground of Analytical Mechanics (canonical transformations, Noether's theorem, etc.).

Another direction which is not considered herein ([3]) is 2D free-surface flows with vorticity, although there has been a lot of progress in this direction in the last two decades. This progress has mainly been based on the existence of a scalar stream function, which permits a drastic simplification of the formulation, not available in three dimensions.

## 2. Differential formulation of the problem

### 2.1. *Generalities. Description of the fluid domain*

In this work, our attention is focused on an inviscid, compressible (barotropic) fluid, undergoing a rotational flow in a horizontally unbounded domain. The fluid domain is limited by a free surface (upper boundary), an impermeable moving bottom (seabed, lower boundary), and -possibly- vertical lateral boundaries, restricting the fluid domain horizontally in some directions (horizontal sectors); see Fig. 1. For simplicity, the lateral boundaries, if existing, are assumed to be rigid walls. To give an exact mathematical formulation of this fluid-dynamics problem, an orthogonal Cartesian system $Ox_1x_2z$ is introduced, with $\boldsymbol{x} = (x_1, x_2)$ being the horizontal spatial variables, and $z$ being the vertical variable, pointing upwards, i.e. in the opposite direction with respect to the constant gravity acceleration $\boldsymbol{g}$. The level $z=0$ is taken to coincide with the quiescent free surface. The moving seabed is located at $z=-h(\boldsymbol{x},t)$, where $h(\boldsymbol{x},t)$ is a known depth function of the horizontal variable $\boldsymbol{x}$ and the time $t$, whereas the free surface is described by the equation $z=\eta(\boldsymbol{x},t)$, where $\eta(\boldsymbol{x},t)$ is an (unknown) surface-elevation function. Accordingly, the fluid domain is shaped as (see Fig. 1)

$$V(t) = \{(\boldsymbol{x},z)\in\mathbb{R}^3 : \boldsymbol{x}=(x_1,x_2)\in D\subseteq\mathbb{R}^2,\ z\in[-h(\boldsymbol{x},t),\eta(\boldsymbol{x},t)]\}, \tag{1}$$

where the horizontal domain $D$ is the projection of the free surface on the plane $(x_1,x_2)$. $D$ is assumed to be unbounded and simply connected.

([3]) With the exception of the two recent papers [27, 28], mentioned above.

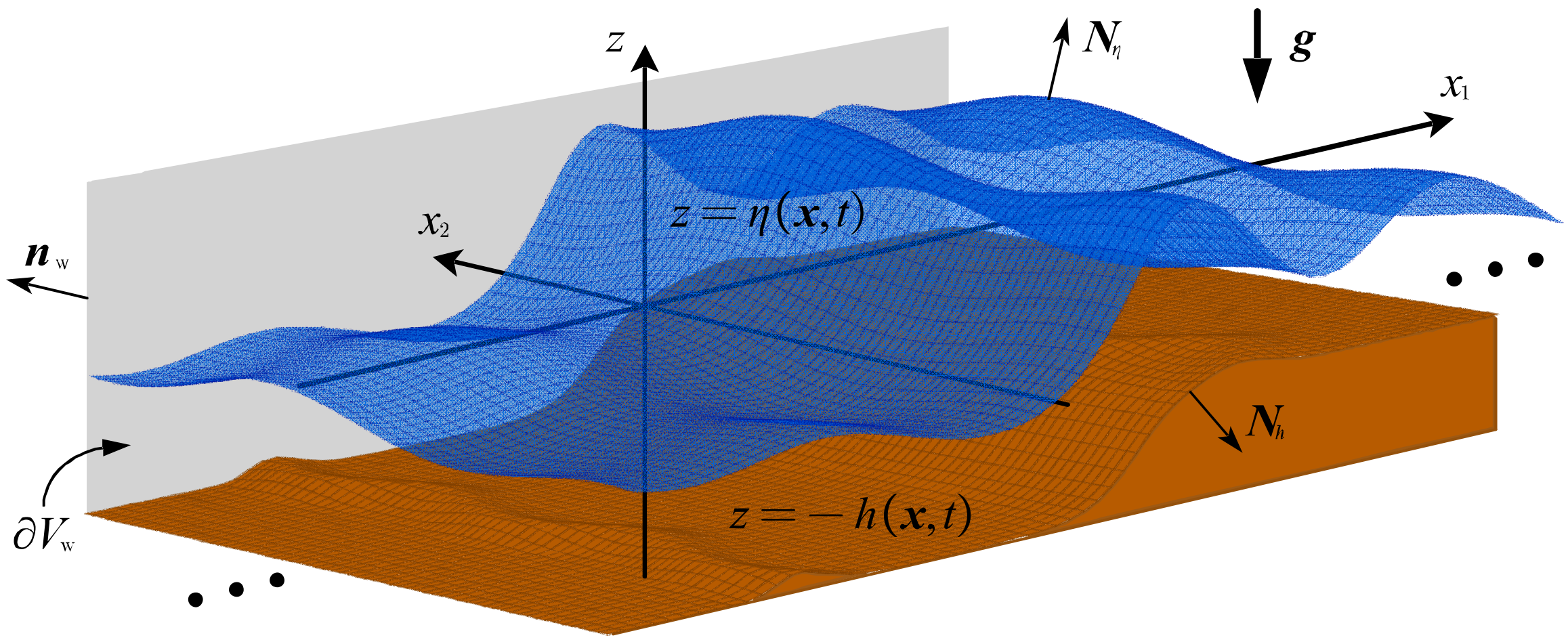

**Figure 1.** Geometrical configuration of the fluid domain $V(t)$.

The vectors

$$\boldsymbol{N}_\eta = \nabla\big(z-\eta(\boldsymbol{x},t)\big) = \left(-\frac{\partial\eta}{\partial x_1},\ -\frac{\partial\eta}{\partial x_2},\ 1\right) \tag{2a}$$

and

$$\boldsymbol{N}_h = -\nabla\big(z+h(\boldsymbol{x},t)\big) = \left(-\frac{\partial h}{\partial x_1},\ -\frac{\partial h}{\partial x_2},\ -1\right) \tag{2b}$$

are normal (perpendicular), but non-unitary, on the free surface and the seabed, respectively, pointing outwardly with respect to the fluid. They commonly appear in our subsequent calculations, and this is why they are given special names.

For convenience, the free surface and the seabed are also denoted by $\partial V_f$ and $\partial V_{sb}$, respectively. The lateral boundary of the fluid domain $V(t)$ is assumed to be vertical, and, as a whole, is denoted by $\partial V_{\mathrm{lat}}$. It consists of two types of boundaries: rigid-wall boundaries, denoted by $\partial V_{\mathrm{w}}$, and infinite "boundaries", meaning that the flow extends to infinity in the corresponding horizontal directions, denoted by $\partial V_\infty$. The outward unit normal vector on the lateral boundary is denoted by $\boldsymbol{n}_{\mathrm{lat}}$, or $\boldsymbol{n}_{\mathrm{w}}$ when the considerations are restricted on the rigid-wall part.

Of the two descriptions normally considered in hydrodynamics, the Eulerian one is of primary interest to us, although the Lagrangian description will also be touched upon, since it is intimately related with Hamilton's Variational Principle. Although our main goal herein is to provide a variational formulation of rotational free-surface flows, in this section we shall briefly present the classical differential formulation of the problem. There are good reasons for doing this: i) Rederiving standard differential equations and boundary conditions from the variational principle is a desirable justification of both approaches, ii) Some of these equations, especially those expressing the kinematics of the flow, will be used as constraints in the variational principle, iii) An interesting, unexpected feature appears in the variational derivation of kinematic boundary conditions. Two types of them are produced variationally: the standard ones, as given in the present section, and an alternative type, consisting of zero tangential velocity. This phenomenon is further commented in Remark 6, at the end of Sec. 5.

### 2.2. *Equations governing the bulk flow*

The involved fields in the Eulerian description are: the fluid velocity $\boldsymbol{u} = \boldsymbol{u}(\boldsymbol{x}, z, t)$, the fluid density $\rho = \rho(\boldsymbol{x}, z, t)$, and the fluid pressure $p = p(\boldsymbol{x}, z, t)$, which, along with the free-surface elevation $\eta = \eta(\boldsymbol{x}, t)$, constitute the unknowns of the problem. The differential equations, governing the fluid flow in $V(t)$, are the following (see e.g. [33–35]):

$$\frac{\partial \rho}{\partial t} + \nabla \cdot (\rho \boldsymbol{u}) = 0, \qquad \text{(conservation of mass)}, \tag{3}$$

$$\frac{\partial \boldsymbol{u}}{\partial t} + (\boldsymbol{u} \cdot \nabla)\boldsymbol{u} + \frac{\nabla p}{\rho} = -g, \qquad \text{(Euler equation)}, \tag{4}$$

which expresses the momentum law, and the constitutive equation (barotropic fluid)

$$p = p(\rho) = \rho^2 \frac{\partial E(\rho)}{\partial \rho}, \tag{5}$$

where $E = E(\rho)$ is the internal energy per unit mass of the fluid (defined by phenomenological considerations). Note that, in Eqs. (3) and (4), $\nabla \bullet = \left( \frac{\partial \bullet}{\partial x_1}, \frac{\partial \bullet}{\partial x_2}, \frac{\partial \bullet}{\partial z} \right)$ is the 3D gradient operator.

**Remark 1:** In many relevant works, the internal energy is assumed to depend on the density, $\rho$, and the specific entropy, $s$, of the fluid, i.e. $E = E(\rho, s)$, in accordance with general thermodynamic considerations. We do not follow this tradition for various reasons. First, the related necessary assumption of isentropic flow cancels any essential contribution from the introduction of the entropy in the variational analysis, since it produces only irrotational flows; see e.g. [10] (p. 5). Second, the option of studying non-isentropic flows (which induce heat transfer) based on classical Hamilton's Principle is controversial [9] (p. 345). Third, the introduction of Lin's constraint, expressing the conservation of parcels' identity (see Sec. 3.1, 3.2), resolves the problem of variationally obtaining the equations of rotational flow, initially supposed to be resolved by using the specific entropy as an independent field.

### 2.3. *Boundary conditions*

As concerns the boundary conditions, we have to distinguish the various types of boundaries. On the free surface, the following two conditions apply

$$\frac{\partial \eta}{\partial t} + u_1 \frac{\partial \eta}{\partial x_1} + u_2 \frac{\partial \eta}{\partial x_2} - u_3 = 0, \qquad \text{on } z = \eta(\boldsymbol{x},t), \tag{6}$$

$$p = \overline{p} = \text{given}, \qquad \text{on } z = \eta(\boldsymbol{x},t), \tag{7}$$

where $\overline{p} = \overline{p}(\boldsymbol{x},t)$ is a known applied-pressure field. The first of them is the kinematic condition, ensuring that the free surface moves in accordance with the motion of the fluid parcels lying on it (that is, it is a material surface), while the second one is of dynamic nature, and it is called the dynamic free-surface condition.

**Remark 2:** In the case of barotropic, irrotational flow, Bernoulli's equation ([4]) permits us to write the dynamic condition (7) as an evolution equation with respect to the velocity potential, greatly facilitating the variational formulation, the mathematical study, and the numerical treatment of the problem [3, 36–40]. Such an equation is not a priori available for the problem under consideration. However, an extended version of Bernoulli's equation, appropriate for the examined problem (rotational flows), can be derived from the variational formulation presented herein ([5]); see Sec. 4.4.

On the moving seabed and on the lateral rigid-wall boundaries, we have the kinematic conditions

$$\frac{\partial h}{\partial t} + u_1 \frac{\partial h}{\partial x_1} + u_2 \frac{\partial h}{\partial x_2} + u_3 = 0, \qquad \text{on } z = -h(\boldsymbol{x},t), \tag{8}$$

and

$$\boldsymbol{u} \cdot \boldsymbol{n}_{\mathrm{w}} = 0, \qquad \text{on } \partial V_{\mathrm{w}}, \tag{9}$$

ensuring zero mass flow through the impermeable boundaries. Recall that, in Eq. (9), $\boldsymbol{n}_{\mathrm{w}}$ is the normal unit vector on the rigid-wall boundary $\partial V_{\mathrm{w}}$. Since the motion of these boundaries is predetermined, a second (dynamical) condition is not needed.

Finally, as concerns the lateral "infinite" boundary $\partial V_{\infty}$, that is, the flow at infinity in the unbounded horizontal directions, we assume that the velocity field and the free-surface elevation tend to zero with rates ensuring that the energy integral is finite. In the opposite case, that is, when a wave system of infinite energy has been developed and moves towards infinity, the variational method is not directly applicable. Nevertheless, such cases can be included in the present formulation by using domain decomposition techniques. Such kinds of techniques have been well developed in the context of irrotational free-surface flows; see e.g. [41]–[43] and [44] (Ch. 7).

---

([4]) Bernoulli's equation for barotropic, irrotational flows takes the form $\frac{p}{\rho} + \frac{\partial \Phi}{\partial t} + \frac{1}{2}\boldsymbol{u}^2 + E + \Omega = C(t)$ [33] (Article 20), where $\Phi$ is the velocity potential ($\boldsymbol{u} = \nabla \Phi$), $\Omega = \Omega(\boldsymbol{x}, z)$ is the potential of an external conservative force-field, and $C(t)$ is an arbitrary function of time.

([5]) Bernoulli-like equations, for some cases of rotational flows, appear (although rarely) in the literature. See e.g. [29] and [33] (Article 167). These equations require the use of Clebsch potentials, which will be obtained variationally later. See Sec. 4.3 and 4.4.

### 2.4. *A brief discussion of the Lagrangian approach*

Analytical mechanics is based on the consideration of material particles' positions $\boldsymbol{x}_i(t)$ and their virtual variations $\delta\boldsymbol{x}_i(t)$, subjected to all relevant constraints. For a continuum, especially a fluid, the index identifying the positions of material parcels is a vector field, leading to an expression of the form $\boldsymbol{x}_{\boldsymbol{a}}(t) = \boldsymbol{x}(\boldsymbol{a},t)$. The common choice of index (label) $\boldsymbol{a}$ is the initial position of the fluid parcels, which means that $\boldsymbol{x}(\boldsymbol{a},t_0) = \boldsymbol{a}$. Clearly, in this consideration, $\boldsymbol{a}$ is an independent variable extending over the geometric domain $V(t_0)$, not varying with time. For each $t \geq t_0$, the transformation

$$V(t_0) \ni \boldsymbol{a} \longrightarrow \boldsymbol{x}(\boldsymbol{a},t) \in V(t) \tag{10a}$$

is assumed regular, smooth and invertible (diffeomorphism). Thus, the inverse transformation,

$$V(t) \ni \boldsymbol{x} \longrightarrow \boldsymbol{a}(\boldsymbol{x},t) \in V(t_0), \tag{10b}$$

is well defined, giving $\boldsymbol{a}$ a representation as a function of $(\boldsymbol{x},t)$. This means that, to a fluid parcel lying at the position $\boldsymbol{x}$ in the time instant $t$, corresponds an initial position $\boldsymbol{a} = \boldsymbol{a}(\boldsymbol{x},t)$. Accordingly, the same $\boldsymbol{a}$ – value is assigned to all $(\boldsymbol{x},t)$ shaping a fluid-parcel trajectory. This means that the field $\boldsymbol{a}(\boldsymbol{x},t)$ is invariant along trajectories, a fact equivalent to the equation

$$\frac{D\boldsymbol{a}(\boldsymbol{x},z,t)}{Dt} \equiv \frac{\partial \boldsymbol{a}}{\partial t} + (\boldsymbol{u}\cdot\nabla)\boldsymbol{a} = 0. \tag{11}$$

Eq. (11), usually referred to as the *conservation of identity*, will be used as a kinematical constraint in the Hamiltonian action functional. Equations for the conservation of mass and the conservation of momentum in terms of the parcel-position field $\boldsymbol{x}(\boldsymbol{a},t)$ can be found in most books of Hydrodynamics (see e.g. [33, 45]). They are not presented herein because they are not needed in our study.

**Remark 3:** All the fields involved in our considerations, both geometrical, as $\eta(\boldsymbol{x},t)$ and $h(\boldsymbol{x},t)$, and physical, as $\rho(\boldsymbol{x},z,t)$, $\boldsymbol{u}(\boldsymbol{x},z,t)$ etc., are considered to be sufficiently smooth. In particular, for the most part of our analysis, $C^1$ – smoothness suffices.

## 3. Variational formulation of the problem

### 3.1. *Preliminary remarks. The variational controversy*

As it is well known, the dynamical equations of any ideal (non-dissipative) mechanical system, subjected to external loads, can be obtained by Hamilton's Principle (see, e.g., [46–48])

$$\delta \int_{\mathrm{T}} \mathcal{L}\, dt + \int_{\mathrm{T}} \delta W\, dt = 0, \tag{12}$$

where $\mathcal{L}$ is the Lagrangian function of the system, $\delta W$ denotes the external loads' virtual work, and $T = [t_0, t_1]$ is an arbitrary time interval. The fundamental (primitive) Lagrangian function is the kinetic minus potential energy, augmented by appropriate, case-dependent, constraints. The virtual work $\delta W$ does not include the contribution of conservative external loads, which are included in the (potential energy of the) Lagrangian via an appropriate force-potential function. The fluid-flow problem examined herein belongs to the class of problems governed by Eq. (12), from which we should be able to derive all the governing equations and boundary conditions, with the proviso that it is correctly implemented. The latter statement means that the variations have to be performed in accordance with the assumptions underlying Hamilton's Principle.

In the Eulerian description of fluid flow, Eqs. (3) – (9), the fluid-parcel positions $\boldsymbol{X}(\boldsymbol{a},t)$ do not come into play and, consequently, the information concerning the parcels' individuality and virtual variations of their positions is apparently lost. This fact is the source of the *variational controversy* of Eulerian fluid dynamics, which has also been discussed in the Introduction. This variational controversy has been solved, at least for the bulk motion of the fluid inside $V(t)$, by means of the clever proposal of Lin [11, 12], to introduce Eq. (11), along with the conservation of mass, Eq. (3), as constraints in the Eulerian action functional. This miraculous trick, whose physical interpretation needs more clarification (at least, to authors' opinion), results in an augmented action functional (see Eq. (16), in Sec. 3.2) that contains both Eulerian and Lagrangian elements. Such a mixed formulation seems to be unpleasant and difficult-to-use, at first glance. However, this difficulty is greatly relaxed, since $\boldsymbol{a}(\boldsymbol{x},z,t)$, and the corresponding Lagrange multiplier $\boldsymbol{A}(\boldsymbol{x},z,t)$, gain a new role, through the Euler-Lagrange equations of the variational principle, as elements of a representation of the velocity and pressure fields, in their Eulerian formulation; see Eqs (32) – (33). These representations are generalizations of similar ones given by Clebsch [29] for incompressible flows. The variational reconstruction of Clebsch representation has been touched upon by Bateman [30, 31], for 2D compressible flows, and has been derived for general 3D flows by Serrin [11]. The issue is discussed in detail by Seliger and Whitham, in their seminal paper [10]; see also [49].

Nice as this phenomenon might be, it seems that it does not resolve the variational controversy on the boundaries of the fluid domain. To resolve this problem, we introduce differential-variational constraints, which allow us to express the Eulerian variations $\delta\boldsymbol{u}(\boldsymbol{x},z,t)$ and $\delta\rho(\boldsymbol{x},z,t)$, on the boundaries, in terms of the Lagrangian ones, resulting in correct kinematic and dynamic boundary conditions. This is the main original contribution of the present work, which is further developed in Sec. 3.3 and 5.

### 3.2. *The Hamiltonian action functional. Bulk and boundary terms*

Since the fluid-flow problem examined herein refers to an ideal mechanical system, its dynamics must be governed by Hamilton's Principle ([6]). This fact was recognized in the early 20th century (see e.g. [8, 9, 50]), and it has been the starting point of many relevant investigations since then. According to this line of thought, the basic (primitive) Lagrangian function is defined as the kinetic minus potential energy of the fluid, that is,

$$\mathcal{L}_{\text{prim}} = \int_{V(t)} \rho\left(\frac{\boldsymbol{u}^2}{2} - E(\rho) - P\right) dV\,, \tag{13}$$

where $P = P(\boldsymbol{x},z)$ is the gravity-force potential. Hamilton's Principle, based on the above Lagrangian (augmented with the mass conservation), produces the dynamics of rotational flows in the Lagrangian description, where the variational controversy does not exist [8, 10]. In 1955

([6]) As always in this paper, we principally refer to the Eulerian formalism of hydrodynamics. For the equations of motion in the Lagrangian description, the connection with Analytical Mechanics is much clearer. Even Lagrange himself gave a derivation of the latter equations by means of D' Alembert's principle of virtual work.

Herivel [9] tried to obtain similar results in the Eulerian description of the flow. His try was unsuccessful, due to the variational controversy, which was not clearly understood at the time. The problem was remedied some years later, after Lin's proposal to add the conservation of identity as a second constraint, apart from the conservation of mass. These observations led to the following, augmented Lagrangian function, which appears in all relevant works after 1959, e.g. [10, 11, 13, 17],

$$\mathcal{L} = \int_{V(t)} \left\{ \rho \left( \frac{\boldsymbol{u}^2}{2} - E(\rho) - P \right) - k \left( \frac{\partial \rho}{\partial t} + \nabla \cdot (\rho \boldsymbol{u}) \right) - \rho \boldsymbol{A} \frac{D\boldsymbol{a}}{Dt} \right\} dV, \tag{14}$$

where $k = k(\boldsymbol{x}, z, t)$ and $\boldsymbol{A} = \boldsymbol{A}(\boldsymbol{x}, z, t) = (A_1, A_2, A_3)$ are the Lagrange multipliers of the mass and identity constraints, respectively.

What remains to completely specify a variational principle of the form (12), for our fluid-flow problem, is the virtual work of the external loads acting on the boundaries of the fluid domain. It turns out that it suffices to specify only the virtual work of loads acting on those parts of the boundary for which a dynamic boundary condition is applied. Since rigid-wall lateral boundaries are assumed fixed, and since the seabed may undergo only predetermined (known) motion, only the free surface belongs to this category. The external load on the free surface is realized by means of an applied pressure $\overline{p}(\boldsymbol{x}, t)$, and the corresponding virtual work is given by the formula

$$\delta W\big|_{\partial V_f} = - \int_D \overline{p}\, \delta\eta \, dx_1\, dx_2 = - \delta \int_D \overline{p}\, \eta \, dx_1\, dx_2 , \tag{15}$$

where $\delta\eta$ is the virtual variation of the free-surface elevation. Note that Eq. (15) is exactly the same as in the irrotational case.

Combining the Lagrangian function (14), and the virtual work (15), we are now in a position to state the complete, augmented action functional for our problem:

$$\begin{aligned} &\tilde{\mathscr{S}}[\boldsymbol{a}, \boldsymbol{u}, \rho, \boldsymbol{A}, k, \eta] = \\ &= \int_{\mathrm{T}} \int_D \int_{-h(\boldsymbol{x},t)}^{\eta(\boldsymbol{x},t)} \left\{ \rho(\cdots) \left( \frac{\boldsymbol{u}^2}{2}(\cdots) - E\big(\rho(\cdots)\big) - P(\boldsymbol{x}, z) \right) \right. \\ &\qquad \left. - k(\cdots) \left( \frac{\partial \rho(\cdots)}{\partial t} + \nabla \cdot \big( \rho(\cdots)\, \boldsymbol{u}(\cdots) \big) \right) - \rho(\cdots) \boldsymbol{A}(\cdots) \frac{D\boldsymbol{a}(\cdots)}{Dt} \right\} dz\, d\boldsymbol{x}\, dt \\ &\qquad - \int_{\mathrm{T}} \int_D \overline{p}(\boldsymbol{x}, t)\, \eta(\boldsymbol{x}, t)\, d\boldsymbol{x}\, dt, \end{aligned} \tag{16}$$

where $(\cdots)$ stands for $(\boldsymbol{x}, z, t)$, and, as in Eq. (12), $T = [t_0, t_1]$ is an arbitrary time interval. Then, the global variational equation, governing the dynamics of the studied problem, takes the form

$$\delta \tilde{\mathscr{S}}[\boldsymbol{a}, \boldsymbol{u}, \rho, \boldsymbol{A}, k, \eta] = 0 . \tag{17}$$

### 3.3. *Differential-variational constraints and boundary virtual displacements*

Although Lin's constraint resolved the variational controversy within the 3D fluid domain, permitting us to consider the Eulerian variations $\delta\rho(\boldsymbol{x},z,t)$, $\delta\boldsymbol{u}(\boldsymbol{x},z,t)$ etc. as independent within $V(t)$, it seems that the problem remains unsolved on the boundary of the fluid domain. Conceptual, a priori, arguments, supporting this statement, are the following:

i) Lin's constraint is an integral constraint acting within the 3D fluid domain $V(t)$. Thus, there is no strong reason to believe that such a constraint works equally well on the boundary, which is a lower-dimension manifold.

ii) There is no evidence that Lin's constraint can handle the variational controversy on moving boundaries, such as the free surface and the moving seabed.

A posteriori justification of the conjecture that Lin's constraint is inadequate for the correct treatment of dynamic boundary conditions, is the fact that we obtain correct conditions using additional, differential-variational constraints (see below). On the contrary, if we treat the Eulerian variations $\delta\rho(\boldsymbol{x},z,t)$, $\delta\boldsymbol{u}(\boldsymbol{x},z,t)$ etc. as independent on the boundary, as well, nothing interesting can be obtained. See Sec. 5.1.

The differential-variational constraints are relations between Lagrangian variations (for fixed $\boldsymbol{a},t$) and Eulerian ones (for fixed $\boldsymbol{x},z,t$), for the same field written in Eulerian variables. To explain these relations, we need to consider both the Lagrangian and Eulerian descriptions of the fluid flow, and introduce two variation operators: $\delta_L(\bullet)$, for variations corresponding to fixed $(\boldsymbol{a},t)$, and $\delta(\bullet)$, for variations corresponding to fixed $(\boldsymbol{x},z,t)$. We further recall that, in the Lagrangian description of the flow, parcel trajectories are defined by equations of the form $\boldsymbol{X}=\boldsymbol{X}(\boldsymbol{a},t)=(X_1,X_2,Z)(\boldsymbol{a},t)$, while, in the Eulerian description, labels $\boldsymbol{a}$ can be considered as a spatial field, $\boldsymbol{a}=\boldsymbol{a}(\boldsymbol{x},z,t)$, in the sense discussed in Sec. 2.4. That said, we are in a position to state the relation between the two variation operators, $\delta_L(\bullet)$ and $\delta(\bullet)$:

$$\delta_L(\bullet) = \delta(\bullet)+(\delta_L\boldsymbol{X}\cdot\nabla)(\bullet), \tag{18}$$

for any field of the fluid written in Eulerian independent variables $\boldsymbol{x},z,t$, where $\delta_L\boldsymbol{X}$ is the virtual displacement of the fluid parcels, also written in Eulerian variables, that is, $\delta_L\boldsymbol{X} = \delta_L\boldsymbol{X}\big(\boldsymbol{a}(\boldsymbol{x},z,t),t\big)$. An analytical proof of Eq. (18) is given by Gelfand and Fomin in [51] (Sec. 37) ([7]). A schematic illustration of the main idea behind the derivation is presented in [16]. A graphical derivation of Eq. (18), inspired by Gelfand and Fomin's analysis, can also be found in [48] (Sec. 6.6). The latter is somewhat obscure, using the natural time of the system also as an artificial "time" associated with the variations. See also [52] (p. 18).

Via the mass conservation, $\delta_L\boldsymbol{X}$ induces corresponding variations of the density, $\delta_L\rho$, which can be conveniently expressed as $\delta_L\rho = -\rho(\nabla\cdot\delta_L\boldsymbol{X})$ [16]. Further, since $\boldsymbol{a}$ are parcel labels and the velocity is the (total) time derivative of the position in the Lagrangian setting, it follows that $\delta_L\boldsymbol{a}=0$ and $\delta_L\boldsymbol{u}=\delta_L\big(D\boldsymbol{X}/Dt\big)$, respectively. Using those relations along with Eq. (18), we conclude that the following *differential-variational constraints* are implied for the Eulerian variations:

([7]) Gelfand and Fomin prove Eq. (18) in the abstract setting of Calculus of Variations. No direct mention is given to Lagrangian and Eulerian Hydrodynamics.

$$\delta \rho = -\nabla \cdot (\rho \, \delta_L \boldsymbol{X}), \tag{19a}$$

$$\delta \boldsymbol{a} = -(\delta_L \boldsymbol{X} \cdot \nabla) \boldsymbol{a}, \tag{19b}$$

and

$$\delta \boldsymbol{u} = \frac{D}{Dt}(\delta_L \boldsymbol{X}) - (\delta_L \boldsymbol{X} \cdot \nabla) \boldsymbol{u}. \tag{19c}$$

Eqs. (19a) and (19b) are the differential-variational counterparts of the mass and identity conservations, while Eq. (19c) is a relation between the variation of the velocity as an independent Eulerian field and its variation as the time derivative of the parcels' trajectories.

The following features of Eqs. (19) are worthwhile to notice:

- Eqs. (19) are point-wise conditions applying to any fluid parcel, lying either in the interior or on the boundary of the fluid domain $V(t)$.
- Using Eqs. (19), the variational equation of the action functional can be re-expressed in terms only of $\delta_L \boldsymbol{X}$.
- If Eqs. (19) are used in the interior points of the fluid domain, the integral constraints appearing in Eq. (16) become redundant. Using such an approach, Bretherton [16] (Sec. 3), directly derived the standard Euler equations, without producing the Clebsch representation for the velocity and pressure fields. We are not following this approach herein, since one of our main goals is to exploit Clebsch representation in the derivation of the free-surface dynamic boundary condition.
- Eqs. (19), when applied to boundary points, should be combined with any additional constraints on the virtual displacements $\delta_L \boldsymbol{X}$, implied by the geometry and the motion of the boundaries.

The last point needs some more elaboration. First, we introduce the notation $\delta_L \boldsymbol{X}_\eta$, $\delta_L \boldsymbol{X}_h$ and $\delta_L \boldsymbol{X}_{\mathrm{w}}$, for the virtual displacements of the fluid parcels lying on the free surface, the seabed, and the lateral rigid-wall boundary, respectively. The synoptic notation $\delta_L \boldsymbol{X}_b$, $b \in \{\eta, h, \mathrm{w}\}$, will be also used, to facilitate general statements concerning boundary virtual displacements. Then, we observe that, in $\delta_L \boldsymbol{X}_b = \delta_L \boldsymbol{X}_b(\boldsymbol{x}, z, t)$, the arguments $(\boldsymbol{x}, z, t)$ are not independent, but should, instead, satisfy the equations defining the corresponding boundary. For example,

$S_\eta(\boldsymbol{x}, z, t) \equiv z - \eta(\boldsymbol{x}, t) = 0$ on the free surface, and

$S_h(\boldsymbol{x}, z, t) \equiv -z - h(\boldsymbol{x}, t) = 0$ on the seabed.

Finally, we state the following conditions for $\delta_L \boldsymbol{X}_b$, dictated by the geometry and kinematics of the corresponding boundaries, in conjunction with the defining properties of the virtual displacements:

On the free surface, $\partial V_f$, we have that

$$\delta_L \boldsymbol{X}_\eta \cdot \boldsymbol{N}_\eta = \delta \eta \qquad (\boldsymbol{x}, z, t) \in \partial V_f(t). \tag{20}$$

Eq. (20) is proved as follows. Since $\partial V_f$ is a freely moving, unknown boundary, $\delta_L \boldsymbol{X}_\eta$ are arbitrary on it. However, $\partial V_f$ is defined by the unknown surface-elevation field $\eta = \eta(\boldsymbol{x}, t)$ and, thus, its variation $\delta \eta$ must be related with $\delta_L \boldsymbol{X}_\eta$. Since the free surface is a material surface, Eq. (18) leads to $0 = \delta_L S_\eta = \delta S_\eta + \delta_L \boldsymbol{X}_\eta \cdot \nabla S_\eta$. From this equation, along with the relations $\delta S_\eta = -\delta\eta$ and $\nabla S_\eta = \boldsymbol{N}_\eta$ (see also Eq. (2a)), we obtain Eq. (20).

Working similarly, we find the following equation for the virtual displacement on the moving seabed,

$$\delta_L \boldsymbol{X}_h \cdot \boldsymbol{N}_h = 0, \qquad (\boldsymbol{x}, z, t) \in \partial V_{\mathrm{sb}}(t), \tag{21}$$

since the motion of this part of the boundary is predetermined. Here, $\boldsymbol{N}_h = \nabla S_h$ (see, also, Eq. (2b)).

Finally, on the lateral rigid-wall boundary, $\partial V_{\mathrm{w}}$, the variation should be compatible with the impermeability condition, which leads to

$$\delta_L \boldsymbol{X}_{\mathrm{w}} \cdot \boldsymbol{n}_{\mathrm{w}} = 0. \qquad (\boldsymbol{x}, z) \in \partial V_{\mathrm{w}}. \tag{22}$$

### 3.4. *Methodology of the variational procedure*

As explained in the previous subsection, in order to cope with the variational controversy arising from the Eulerian description of the involved fields, the volume integral terms, appearing in the variational equation, will be treated differently than the boundary integral terms ([8]). In the present subsection, we briefly describe the main steps taken in Sec. 4 and 5, towards the implementation of this, somewhat unconventional, approach.

**Step 1:** First, we calculate the partial Gateaux derivatives

$$\delta_q \tilde{\mathscr{S}}[\boldsymbol{a}, \rho, \boldsymbol{A}, k, \boldsymbol{u}, \eta; \delta q], \qquad q \in \{\boldsymbol{a}, \rho, \boldsymbol{A}, k, \boldsymbol{u}, \eta\}, \tag{23}$$

of the augmented action functional (16), and write the global variational Eq. (17) in the expanded form

$$\underbrace{\sum_q \int_{\mathrm{T}} \int_D \int_{-h(\boldsymbol{x},t)}^{\eta(\boldsymbol{x},t)} (\cdots)\, \delta q \, dz \, d\boldsymbol{x} \, dt}_{\text{volume integral terms}}$$

$$+ \underbrace{\sum_q \int_{\mathrm{T}} \int_D \big[(\cdots)\, \delta q\big]_{z=\eta} \, d\boldsymbol{x} \, dt}_{\text{free-surface boundary terms}} + \underbrace{\sum_q \int_{\mathrm{T}} \int_D \big[(\cdots)\, \delta q\big]_{z=-h} \, d\boldsymbol{x} \, dt}_{\text{seabed boundary terms}}$$

$$+ \underbrace{\sum_q \int_{\mathrm{T}} \int_{\partial V_{\mathrm{w}}} (\cdots)\, \delta q \, dS_{\mathrm{w}} \, dt}_{\text{lateral rigid-wall terms}} = 0. \tag{24}$$

In the last sum of (boundary) integrals, appearing in the left-hand side of Eq. (24), the integration is taken over the whole lateral boundary. However, it suffices to keep the integral only on the rigid-wall part $\partial V_{\rm w}$, as shown above, since the variations of the flow fields are taken to vanish on the "infinite" lateral boundary $\partial V_{\infty}$. Also, $dS_{\rm w}$ denotes the surface element of the rigid wall. Step 1 is realized in Sec. 4.1.

([8]) What are called here "volume integral terms" and "boundary integral terms" are, in fact, time-volume and time-boundary integrals. Nevertheless, here, and subsequently, we do not mention the time integration, since our purpose is to just distinguish the spatially 3D from the spatially 2D terms.

**Step 2:** Then, we consider variations that vanish on the boundaries, and obtain the individual variational equations (since $\delta q$ are considered independent)

$$\int_{\rm T}\int_{D}\int_{-h(\boldsymbol{x},t)}^{\eta(\boldsymbol{x},t)} (\cdots)\,\delta q\, dz\, d\boldsymbol{x}\, dt = 0, \qquad q \in \{\boldsymbol{a},\rho,\boldsymbol{A},k,\boldsymbol{u}\}. \text{ (}^9\text{)} \tag{25}$$

These equations, in conjunction with the fact that $\delta q$ are arbitrary within the fluid domain, provide us with five Euler-Lagrange equations (see Sec. 4.2), which are further discussed in Sec. 4.3 and 4.4. The equations obtained in this step are well known, so that the first two steps do not produce original results. They are, however, necessary prerequisites for the next step, where the original contribution of this paper lies.

**Step 3:** This is taken in Sec. 5. Substituting the Euler-Lagrange equations into the global variational Eq. (17), the volume integral terms are eliminated and there remain only the boundary integral terms, associated with the free surface, the seabed, and the lateral boundaries. As discussed in Sec. 3.3, the boundary variations of the involved Eulerian fields, namely, $\delta\rho$, $\delta\boldsymbol{a}$, $\delta\boldsymbol{u}$, $\delta\eta$, cannot be considered as independent there ([10]), but may all be expressed in terms of the boundary parcels' virtual displacements $\delta_L \boldsymbol{X}_b = \delta_L \boldsymbol{X}_b(\boldsymbol{x},z,t)$, by using Eqs. (19) – (20). Thus, the global variational Eq. (17) reduces to one restricted on the boundary of the fluid domain, involving the variations $\delta_L \boldsymbol{X}_\eta$, $\delta_L \boldsymbol{X}_h$, $\delta_L \boldsymbol{X}_{\rm w}$. The latter equation, in conjunction with Eqs. (21) – (22), and the standard variational arguments, provides us with all (kinematic and dynamic) boundary conditions for the three kinds of boundaries of the studied problem. To the best of our knowledge, this set of boundary conditions is variationally derived for the first time.

## 4. Calculation of variations, and Euler-Lagrange equations within the fluid domain

### 4.1. *Partial Gateaux derivatives of the action functional*

Now, we proceed with the implementation of the methodology described in Sec. 3.4. The starting point is the calculation of partial Gateaux derivatives of the augmented action functional, Eq. (16), with respect to all its arguments. The Gateaux derivatives with respect to the Lagrange multipliers $k$ and $\boldsymbol{A}$ are trivially calculated, resulting in

$$\delta_k \tilde{\mathscr{S}} = -\int_{\rm T}\int_{D}\int_{-h(\boldsymbol{x},t)}^{\eta(\boldsymbol{x},t)} \left(\frac{\partial\rho}{\partial t} + \nabla\cdot(\rho\,\boldsymbol{u})\right)\delta k\, dz\, d\boldsymbol{x}\, dt, \tag{26a}$$

and

$$\delta_A \tilde{\mathscr{S}} = -\int_{\rm T}\int_{D}\int_{-h(\boldsymbol{x},t)}^{\eta(\boldsymbol{x},t)} \rho\frac{D\boldsymbol{a}}{Dt}\delta\boldsymbol{A}\, dz\, d\boldsymbol{x}\, dt. \tag{26b}$$

---

([9]) The variation $\delta\eta$ does not appear in the volume integral terms, since it is related only with the free-surface boundary term.
([10]) Because of the variational controversy.

The three Gateaux derivatives, with respect to the fields $\boldsymbol{a}$, $\boldsymbol{A}$ and $\boldsymbol{u}$, are much more involved, requiring extensive calculations. The corresponding results, in a form appropriate for the subsequent variational analysis, are as follows:

$$\delta_{\boldsymbol{a}} \tilde{\mathscr{S}} = \int_{\mathrm{T}}\int_{D}\int_{-h(\boldsymbol{x},t)}^{\eta(\boldsymbol{x},t)} \left( \frac{D(\rho \boldsymbol{A})}{Dt} + (\nabla\cdot\boldsymbol{u})\,\rho\boldsymbol{A}\right)\delta\boldsymbol{a}\, dz\, d\boldsymbol{x}\, dt$$
$$+ \int_{\mathrm{T}}\int_{D} \left(\frac{\partial\eta}{\partial t} - [\boldsymbol{u}]_{z=\eta}\,\boldsymbol{N}_\eta\right)\left[\rho\,\boldsymbol{A}\,\delta\boldsymbol{a}\right]_{z=\eta} d\boldsymbol{x}\, dt \tag{26c}$$
$$+ \int_{\mathrm{T}}\int_{D} \left(\frac{\partial h}{\partial t} - [\boldsymbol{u}]_{z=-h}\,\boldsymbol{N}_h\right)\left[\rho\,\boldsymbol{A}\,\delta\boldsymbol{a}\right]_{z=-h} d\boldsymbol{x}\, dt$$
$$- \int_{\mathrm{T}}\int_{\partial D_{\mathrm{w}}}\int_{-h(\boldsymbol{x},t)}^{\eta(\boldsymbol{x},t)} (\rho\boldsymbol{A}\,\delta\boldsymbol{a})\,\boldsymbol{u}\,\boldsymbol{n}_{\mathrm{w}}\, dz\, dl\, dt,$$

$$\delta_{\rho} \tilde{\mathscr{S}} = \int_{\mathrm{T}}\int_{D}\int_{-h(\boldsymbol{x},t)}^{\eta(\boldsymbol{x},t)} \left( \frac{\boldsymbol{u}^2}{2} - E - \rho\frac{\partial E}{\partial\rho} - P + \frac{Dk}{Dt} - \boldsymbol{A}\frac{D\boldsymbol{a}}{Dt}\right)\delta\rho\, dz\, d\boldsymbol{x}\, dt$$
$$+ \int_{\mathrm{T}}\int_{D} \left(\frac{\partial\eta}{\partial t} - [\boldsymbol{u}]_{z=\eta}\,\boldsymbol{N}_\eta\right)\left[k\,\delta\rho\right]_{z=\eta} d\boldsymbol{x}\, dt$$
$$+ \int_{\mathrm{T}}\int_{D} \left(\frac{\partial h}{\partial t} - [\boldsymbol{u}]_{z=-h}\,\boldsymbol{N}_h\right)\left[k\,\delta\rho\right]_{z=-h} d\boldsymbol{x}\, dt \tag{26d}$$
$$- \int_{\mathrm{T}}\int_{\partial D_{\mathrm{w}}}\int_{-h(\boldsymbol{x},t)}^{\eta(\boldsymbol{x},t)} (k\,\delta\rho)\,\boldsymbol{u}\,\boldsymbol{n}_{\mathrm{w}}\, dz\, dl\, dt$$

and

$$\delta_{\boldsymbol{u}} \tilde{\mathscr{S}} = \int_{\mathrm{T}}\int_{D}\int_{-h(\boldsymbol{x},t)}^{\eta(\boldsymbol{x},t)} \rho\,(\boldsymbol{u} + \nabla k - \boldsymbol{A}\,\nabla\boldsymbol{a})\,\delta\boldsymbol{u}\, dz\, d\boldsymbol{x}\, dt$$
$$- \int_{\mathrm{T}}\int_{D} \left[\rho\, k\,\delta\boldsymbol{u}\,\boldsymbol{N}_\eta\right]_{z=\eta} d\boldsymbol{x}\, dt$$
$$- \int_{\mathrm{T}}\int_{D} \left[\rho\, k\,\delta\boldsymbol{u}\,\boldsymbol{N}_h\right]_{z=-h} d\boldsymbol{x}\, dt \tag{26e}$$
$$- \int_{\mathrm{T}}\int_{\partial D_{\mathrm{w}}}\int_{-h(\boldsymbol{x},t)}^{\eta(\boldsymbol{x},t)} \rho\, k\,\delta\boldsymbol{u}\,\boldsymbol{n}_{\mathrm{w}}\, dz\, dl\, dt,$$

where (we recall that) $D$ is the projection of the free surface on the horizontal plane, $\boldsymbol{n}_{\mathrm{w}}$ is the outward unit normal vector on the (vertical) rigid wall, and $\boldsymbol{N}_\eta$, $\boldsymbol{N}_h$ are defined by Eqs. (2).

Further, $\partial D_{\mathrm{w}}$ is the intersection of the lateral rigid-wall boundary with $D$ (that is, $\partial D_{\mathrm{w}}$ is a line), and $dl$ is the corresponding line element. ([11])

The first integral in the right-hand side of each of Eqs. (26c,d,e) is a volume integral, over the whole fluid domain $V(t)$, while the second, third and fourth integrals are surface integrals taken over the free surface, the seabed, and the lateral rigid wall, respectively.

It remains to calculate the Gateaux derivative with respect to the free-surface elevation $\eta$. This is easily performed, by invoking the Leibnitz integral rule, resulting in

$$\delta_\eta \tilde{\mathscr{S}} = \int_{\mathrm{T}} \int_{D} \left[ \rho\left(\frac{1}{2}\boldsymbol{u}^2 - E - P\right) - k\left(\frac{\partial \rho}{\partial t} + \nabla\cdot(\rho\,\boldsymbol{u})\right) - \rho\boldsymbol{A}\frac{D\boldsymbol{a}}{Dt} - \overline{p}\right]_{z=\eta} \delta\eta\, d\boldsymbol{x}\, dt\,. \quad (26\text{f})$$

Consequently, the total variation of the action functional (16) takes the form

$$\delta\tilde{\mathscr{S}} = \sum_q \delta_q \tilde{\mathscr{S}}, \qquad q \in \{\boldsymbol{a},\rho,\boldsymbol{A},k,\boldsymbol{u},\eta\}, \quad (27)$$

where $\delta_q \tilde{\mathscr{S}}$ are given by Eqs. (26a) – (26f).

### 4.2. *Euler-Lagrange equations corresponding to variations within the fluid domain*

Having calculated the partial Gateaux derivatives of the action functional, Eqs. (26), we are able to proceed with the second step of Sec. 3.4, considering variations $\delta\boldsymbol{a}, \delta\rho, \delta\boldsymbol{A}, \delta k, \delta\boldsymbol{u}$ that vanish on the boundary of the fluid domain $V(t)$, and $\delta\eta = 0$. Thus, only volume integral terms survive in the global variational equation. Now, recalling that the variations $\delta\boldsymbol{a}(\boldsymbol{x},z,t)$, $\delta\rho(\boldsymbol{x},z,t)$ etc., $(\boldsymbol{x},z)\in V(t)$, can be considered as independent from each other and arbitrary, and using the standard argument of the Calculus of Variations, we obtain:

$$\delta k: \quad \frac{\partial \rho}{\partial t} + \nabla\cdot(\rho\,\boldsymbol{u}) = 0, \qquad (\boldsymbol{x},z)\in V(t), \quad (28)$$

$$\delta\boldsymbol{A}: \quad \frac{D\boldsymbol{a}}{Dt} \equiv \frac{\partial \boldsymbol{a}}{\partial t} + (\boldsymbol{u}\cdot\nabla)\,\boldsymbol{a} = 0, \qquad (\boldsymbol{x},z)\in V(t), \quad (29)$$

$$\delta\boldsymbol{a}: \quad \frac{D(\rho\boldsymbol{A})}{Dt} + (\boldsymbol{u}\cdot\nabla)(\rho\boldsymbol{A}) = 0,$$

which, using Eq. (28), is rewritten as

$$\frac{D\boldsymbol{A}}{Dt} \equiv \frac{\partial \boldsymbol{A}}{\partial t} + (\boldsymbol{u}\cdot\nabla)\,\boldsymbol{A} = 0, \qquad (\boldsymbol{x},z)\in V(t), \quad (30)$$

$$\delta\rho: \quad \frac{Dk}{Dt} = -\frac{\boldsymbol{u}^2}{2} + E + \rho\frac{\partial E}{\partial\rho} + P + \boldsymbol{A}\frac{D\boldsymbol{a}}{Dt}, \qquad (\boldsymbol{x},z)\in V(t), \quad (31)$$

([11]) The main tools for obtaining Eqs. (26c,d,e) are the Leibnitz integral rule and the divergence theorem. Also, the standard assumption (in Hamilton's Principle) that the variations of all fields are taken as zero at the initial and final times, $t_0, t_1$, is used. The detailed calculations can be found in Appendix A.

$$\delta \boldsymbol{u}: \qquad \boldsymbol{u} = -\nabla k + \boldsymbol{A}\nabla \boldsymbol{a} = -\nabla k + \sum_{i=1}^{3} A_i \nabla a_i\,, \quad (\boldsymbol{x},z) \in V(t)\,. \tag{32}$$

Eqs. (28) and (29) express the conservation of mass and the conservation of identity, directly obtained due to the Lagrange multipliers $k$ and $\boldsymbol{A}$. Eq. (30) shows that Lagrange multipliers $\boldsymbol{A}$ are integrals (constants) of the fluid motion. Eq. (31), simplified by setting its last term equal to zero (because of Eq. (29)), provides an evolution equation for the Lagrange multiplier $k$. Finally, Eq. (32) defines a representation of the velocity field in terms of the fields $k, \boldsymbol{a}, \boldsymbol{A}$. These equations have been obtained (sometimes with a slightly different form) by many authors since 1959, first appearing in [11]. Euler's momentum equation is not directly included in the above list; however, it can be derived by combining Eqs. (29) – (32) and eliminating the fields $k, \boldsymbol{a}, \boldsymbol{A}$, which do not appear in the standard Eulerian formalism. The detailed derivation is given in various books or papers, e.g. [11, 13, 49] and [53] (Ch. 3). Inverting the point of view, we may say that, in the present variational formulation, Euler's momentum Eq. (4) is decomposed into the system of Eqs. (29) – (32). In some sense, Eqs. (28) – (32) provide us with a new formulation of the rotational flow problem. This formulation is Eulerian in the sense that all the involved fields depend on the Eulerian variables $(\boldsymbol{x}, z)$, but it also contains the additional fields $k, \boldsymbol{a}, \boldsymbol{A}$, introduced for resolving the variational controversy. The origin of the latter fields is traced back to the Lagrangian description.

### 4.3. *On the representation of the fluid velocity by means of potentials*

Eq. (32) gives to the three (seven scalar) fields $k, \boldsymbol{a}, \boldsymbol{A}$ the role of extended potentials, defining a representation of the velocity field. It is not clear, however, how many pairs $A_i$, $a_i$ should be used in the term $\boldsymbol{A}\nabla\boldsymbol{a} = \sum_i A_i \nabla a_i$. Although in our formulation we consider $\boldsymbol{a}$ and $\boldsymbol{A}$ as 3D fields, everything can be repeated successfully by assuming that $\boldsymbol{a}$ and $\boldsymbol{A}$ are scalar fields ([12]), obtaining, then, a simpler representation of the velocity by means of three scalar potentials. The latter is exactly the same as the representation derived by Clebsch in [29], using different arguments. The question of how many terms should be kept in the representation $\boldsymbol{u} = -\nabla k + \sum A_i \nabla a_i$ is interesting and important, and has been discussed by many authors; see e.g. [16] (Sec. 6), [17, 21, 22] and [26] (Sec. 9.3). The simple choice of Clebsch representation seems reasonable (since it involves only three scalar potentials), but leads to restrictions on the structure of the vorticity field. In fact, there exist several works pointing out that using the classical Clebsch representation is inadequate for a wide class of problems, e.g. knotted vortex filaments, points of vanishing vorticity [16, 21, 22, 25, 54]. The issue is also closely related with the non-uniqueness of potentials representing a given $\boldsymbol{u}$ (gauge transformations), which is well known for many years [14], but has only recently been analyzed in depth [22–24]. A more detailed discussion of this matter is out of the scope of the present work. It is our plan to come back to this issue in the near future.

### 4.4. *On the representation of the fluid pressure by means of potentials*

Apart from the representation of the velocity field, Eq. (32), an explicit representation for the pressure field, in terms of the potentials $k$, $A_i$ and $a_i$, can be easily found out from Eq. (31).

---

([12]) Which means that one uses only one out of the three scalar Lin's constraints, weakening the conditions of the conservation of identity.

Indeed, using Eq. (5), we see that $\rho\,\partial E/\partial\rho = p/\rho$. Combining the latter with Eq. (31), we obtain

$$\frac{p}{\rho} = \frac{Dk}{Dt} - \boldsymbol{A}\frac{D\boldsymbol{a}}{Dt} + \frac{\boldsymbol{u}^2}{2} - E - P. \tag{33a}$$

An alternative representation of the pressure occurs by using again the expression $\rho\,\partial E/\partial\rho = p/\rho$, in conjunction with Eq. (32). Then, after straightforward calculations, we obtain

$$\frac{p}{\rho} = \frac{\partial k}{\partial t} - \boldsymbol{A}\frac{\partial\boldsymbol{a}}{\partial t} - \frac{\boldsymbol{u}^2}{2} - E - P. \tag{33b}$$

The right-hand side of the latter equation, where $\boldsymbol{u} = \boldsymbol{u}(k, A_i, a_i)$, is closely related (although more general) to the Lagrangian density provided in works relevant with the approach of Clebsch [29]; see, also, [55]. Further, Eqs. (33) are essential for the establishment of the connection between Eq. (7) and the two alternative forms of the dynamic free-surface condition that occur variationally in Sec 5; see Eqs. (51) – (52).

## 5. Boundary-variational equation. Derivation of boundary conditions

Having obtained the Euler-Lagrange equations within the fluid domain, that is, those equations corresponding to independent variations of the fields $k, \boldsymbol{A}, \boldsymbol{a}, \rho$ and $\boldsymbol{u}$ in $V(t)$, there remains to treat the boundary terms of the variational equation and deduce the boundary conditions. After substituting Eqs. (28) – (32) into the total variational Eq. (27), the volume-integral terms are eliminated, and what remains is the following *boundary-variational equation*

$$\begin{aligned}
&\int_{\mathrm{T}}\int_{D}\Bigg\{\left[\left(\frac{\partial\eta}{\partial t} - \boldsymbol{u}\,\boldsymbol{N}_\eta\right)(k\,\delta\rho + \rho\,\boldsymbol{A}\,\delta\boldsymbol{a}) - \rho\,k\,\delta\boldsymbol{u}\,\boldsymbol{N}_\eta\right]_{z=\eta} \\
&\qquad + \left[\rho\left(\frac{\boldsymbol{u}^2}{2} - E - P\right) - k\left(\frac{\partial\rho}{\partial t} + \nabla\cdot(\rho\,\boldsymbol{u})\right) - \rho\,\boldsymbol{A}\frac{D\boldsymbol{a}}{Dt} - \overline{p}\right]_{z=\eta}\delta\eta\Bigg\}\,d\boldsymbol{x}\,dt \\
&+ \int_{\mathrm{T}}\int_{D}\left\{\left(\frac{\partial h}{\partial t} - \boldsymbol{u}\,\boldsymbol{N}_h\right)(k\,\delta\rho + \rho\,\boldsymbol{A}\,\delta\boldsymbol{a}) - \rho\,k\,\delta\boldsymbol{u}\,\boldsymbol{N}_h\right\}_{z=-h} d\boldsymbol{x}\,dt \\
&- \int_{\mathrm{T}}\int_{\partial D_{\mathrm{w}}}\int_{-h(\boldsymbol{x},t)}^{\eta(\boldsymbol{x},t)}\left\{(k\,\delta\rho + \rho\boldsymbol{A}\,\delta\boldsymbol{a})\,\boldsymbol{u}\,\boldsymbol{n}_{\mathrm{w}} + \rho\,k\,\delta\boldsymbol{u}\,\boldsymbol{n}_{\mathrm{w}}\right\}dz\,dl\,dt.
\end{aligned} \tag{34}$$

In the above, the first integral is evaluated on the free surface, the second on the seabed and the third on the lateral boundary.

### 5.1. *Transformation of the boundary-variational equation using the differential-variational constraints*

As discussed in Sec. 3.3, the variations of the various fields on the boundary cannot be considered independent from each other. To better understand this statement, it is expedient to observe what happens if one makes the assumption that the variations $\delta\rho$, $\delta\boldsymbol{a}$ and $\delta\boldsymbol{u}$ *are* independent in the right-hand side of Eq. (34). This exercise leads to a disintegration of each boundary term to more (two terms from $\delta\rho$ and $\delta\boldsymbol{u}\,\boldsymbol{n}_b$, plus the number of $\delta a_i$ utilized in the formulation), and results in a repetitive derivation of the kinematic boundary conditions. This disintegration is also responsible for the impossibility of deriving the dynamic boundary condition on the free surface. These observations provide an additional ad hoc justification of our thesis to express the boundary variations in Eq. (34) in terms of the virtual displacements $\delta_L \boldsymbol{X}_b$, $b \in \{\eta, h, \mathrm{w}\}$, on each boundary, using the differential-variational constraints given by Eqs. (19) – (20).

In that process, attention should be paid to the substitution of the variations $\delta\rho$, $\delta\boldsymbol{a}$ and $\delta\boldsymbol{u}$ in the first two integrals of Eq. (34), referring to the free surface and the seabed. These field variations are generally functions of the variables $\boldsymbol{x}$, $z$ and $t$, where, in this instance, the vertical variable is evaluated on $z=\eta$ or $z=-h$. On the other hand, the virtual displacements $\delta_L \boldsymbol{X}_b$, $b \in \{\eta, h\}$, are -by their nature- position variations of the material parcels that form the free-surface and seabed surfaces and, thus, they are independent of $z$. Consequently, the differential-variational constraints (19a) and (19b) on the free surface and the seabed should be reformulated as

$$\begin{aligned}\left[\delta\rho\right]_{z=\{\eta,-h\}} &= -\left[\nabla\rho\right]_{z=\{\eta,-h\}}\cdot\delta_L \boldsymbol{X}_{\{\eta,h\}} - \left[\rho\right]_{z=\{\eta,-h\}}(\nabla\cdot\delta_L \boldsymbol{X}_{\{\eta,h\}}) = \\ &= -\left[\nabla\rho\right]_{z=\{\eta,-h\}}\cdot\delta_L \boldsymbol{X}_{\{\eta,h\}} \\ &\qquad -\left[\rho\right]_{z=\{\eta,-h\}}\left(\nabla_2\cdot(\delta_L X_{\{\eta,h\},1},\delta_L X_{\{\eta,h\},2})\right)\end{aligned} \tag{35}$$

and

$$\left[\delta\boldsymbol{a}\right]_{z=\{\eta,-h\}} = -\left[\nabla\boldsymbol{a}\right]_{z=\{\eta,-h\}}\cdot\delta_L \boldsymbol{X}_{\{\eta,h\}}, \tag{36}$$

where $\nabla_2(\bullet)\equiv(\partial/\partial x_1,\partial/\partial x_2)(\bullet)$ is the 2D-horizontal gradient. Further, using Eq. (19c), the product $\delta\boldsymbol{u}\,\boldsymbol{N}_b$, $b\in\{\eta,h\}$, is written as

$$\left[\delta\boldsymbol{u}\right]_{z=-h}\boldsymbol{N}_h = -\left(\frac{\partial\boldsymbol{N}_h}{\partial t}+\left[\nabla(\boldsymbol{u}\,\boldsymbol{N}_h)\right]_{z=-h}\right)\delta_L\boldsymbol{X}_h, \qquad z=-h, \tag{37}$$

recalling Eq. (21), and

$$\begin{aligned}\left[\delta\boldsymbol{u}\right]_{z=\eta}\boldsymbol{N}_\eta &= \frac{\partial(\delta_L\boldsymbol{X}_\eta\,\boldsymbol{N}_\eta)}{\partial t}+\left[(u_1,u_2)\right]_{z=\eta}\cdot\nabla_2(\delta_L\boldsymbol{X}_\eta\,\boldsymbol{N}_\eta) \\ &\quad -\left(\frac{\partial\boldsymbol{N}_\eta}{\partial t}+\left[\nabla(\boldsymbol{u}\,\boldsymbol{N}_\eta)\right]_{z=\eta}\right)\delta_L\boldsymbol{X}_\eta, \qquad z=\eta.\end{aligned} \tag{38}$$

For the same product on the lateral rigid-wall boundary, we have that

$$\delta\boldsymbol{u}\,\boldsymbol{n}_\mathrm{w} = -\nabla(\boldsymbol{u}\,\boldsymbol{n}_\mathrm{w})\,\delta_L\boldsymbol{X}_\mathrm{w}, \qquad (\boldsymbol{x},z)\in\partial V_\mathrm{w}, \tag{39}$$

as is readily found by using Eq. (22) and the fact that $\boldsymbol{n}_\mathrm{w}$ is independent of time ($\partial V_\mathrm{w}$ is a known, fixed boundary).

Now, substituting the variations $\delta\rho$, $\delta\boldsymbol{a}$, $\delta\boldsymbol{u}$ and $\delta\eta$, in Eq. (34), with appropriate expressions of $\delta_L\boldsymbol{X}_b$, using Eqs. (35) – (39) and Eq. (20), the former equation is rewritten in terms of the variations $\delta_L \boldsymbol{X}_\eta$, $\delta_L \boldsymbol{X}_h$ and $\delta_L \boldsymbol{X}_{\mathrm{w}}$. These three variations are clearly independent. Thus, considering $\delta_L \boldsymbol{X}_\eta =$ arbitrary, $\delta_L \boldsymbol{X}_h = \delta_L \boldsymbol{X}_w = 0$, we obtain the free-surface term

$$\begin{aligned}\int_{\mathrm{T}}\int_{D}\Bigg\{&\left[-\left(\frac{\partial\eta}{\partial t}-\boldsymbol{u}\,\boldsymbol{N}_\eta\right)(k\,\nabla\rho+\rho\,\boldsymbol{A}\nabla\boldsymbol{a})+\rho\,k\left(\frac{\partial\boldsymbol{N}_\eta}{\partial t}+\nabla(\boldsymbol{u}\,\boldsymbol{N}_\eta)\right)\right]_{z=\eta}\delta_L\boldsymbol{X}_\eta\\&+\left[\rho\left(\frac{\boldsymbol{u}^2}{2}-E-P\right)-k\left(\frac{\partial\rho}{\partial t}+\nabla\cdot(\rho\,\boldsymbol{u})\right)-\rho\,\boldsymbol{A}\frac{D\boldsymbol{a}}{Dt}-\bar{p}\right]_{z=\eta}\delta_L\boldsymbol{X}_\eta\,\boldsymbol{N}_\eta\\&-\left[\rho\,k\left(\frac{\partial\eta}{\partial t}-\boldsymbol{u}\,\boldsymbol{N}_\eta\right)\right]_{z=\eta}\nabla_2\cdot(\delta_L X_{\eta,1},\delta_L X_{\eta,2})\\&-\left[\rho\,k\right]_{z=\eta}\frac{\partial}{\partial t}(\delta_L\boldsymbol{X}_\eta\,\boldsymbol{N}_\eta)-\left[\rho\,k\,(u_1,u_2)\right]_{z=\eta}\nabla_2(\delta_L\boldsymbol{X}_\eta\,\boldsymbol{N}_\eta)\Bigg\}\,d\boldsymbol{x}\,dt\;=\;0,\end{aligned}\tag{40}$$

and similarly, we obtain the moving seabed term

$$\begin{aligned}\int_{\mathrm{T}}\int_{D}\Bigg\{&\left[-\left(\frac{\partial h}{\partial t}-\boldsymbol{u}\,\boldsymbol{N}_h\right)(k\,\nabla\rho+\rho\,\boldsymbol{A}\nabla\boldsymbol{a})+\rho\,k\left(\frac{\partial\boldsymbol{N}_h}{\partial t}+\nabla(\boldsymbol{u}\,\boldsymbol{N}_h)\right)\right]_{z=-h}\delta_L\boldsymbol{X}_h\\&-\left[\rho\,k\left(\frac{\partial h}{\partial t}-\boldsymbol{u}\,\boldsymbol{N}_h\right)\right]_{z=-h}\nabla_2\cdot(\delta_L X_{h,1},\delta_L X_{h,2})\Bigg\}\,d\boldsymbol{x}\,dt\;=\;0,\end{aligned}\tag{41}$$

and the lateral rigid wall term

$$\begin{aligned}\int_{\mathrm{T}}\int_{\partial D_{\mathrm{w}}}\int_{-h(\boldsymbol{x},t)}^{\eta(\boldsymbol{x},t)}\Big\{&-\left[(\boldsymbol{u}\,\boldsymbol{n}_{\mathrm{w}})(k\,\nabla\rho+\rho\,\boldsymbol{A}\nabla\boldsymbol{a})+\rho\,k\,\nabla(\boldsymbol{u}\,\boldsymbol{n}_{\mathrm{w}})\right]\delta_L\boldsymbol{X}_{\mathrm{w}}\\&-\rho\,k\,(\boldsymbol{u}\,\boldsymbol{n}_{\mathrm{w}})\,\nabla\cdot(\delta_L\boldsymbol{X}_{\mathrm{w}})\Big\}\,dz\,dl\,dt\;=\;0.\end{aligned}\tag{42}$$

### 5.2. *Decomposition of boundary virtual displacements into normal and tangential components*

To facilitate the further treatment of the variational Eqs. (40)-(42), we analyze each $\delta_L \boldsymbol{X}_b$, $b \in \{\eta, h, \mathrm{w}\}$, into its normal and tangential components; i.e. we write

$$\delta_L \boldsymbol{X}_b \;=\; \delta_L \boldsymbol{X}_{b,\perp} + \delta_L \boldsymbol{X}_{b,\parallel}, \qquad b \in \{\eta, h, \mathrm{w}\}. \tag{43}$$

The normal components $\delta_L \boldsymbol{X}_{b,\perp}$ are expressed as

$$\begin{cases}\delta_L \boldsymbol{X}_{b,\perp} \;=\; \delta B_{b,\perp}\dfrac{\boldsymbol{N}_b}{\left\|\boldsymbol{N}_b\right\|^2}, & b \in \{\eta, h\},\\[2ex] \delta_L \boldsymbol{X}_{\mathrm{w},\perp} \;=\; \delta B_{\mathrm{w},\perp}\,\boldsymbol{n}_{\mathrm{w}}, & \end{cases}\tag{44}$$

where $\delta B_{b,\perp}$ are scalar quantities ([13]). On the free surface and the seabed, the tangential component $\delta_L \boldsymbol{X}_{b,\|}$ is expanded in the natural basis of the tangent planes. Considering the parametric representations of the free surface and the seabed, at frozen time $t = \tilde{t}$ ,

$$\boldsymbol{r}_\eta(\boldsymbol{x}) = \big(x_1, x_2, \eta(x_1, x_2, \tilde{t})\big), \qquad \boldsymbol{r}_h(\boldsymbol{x}) = \big(x_1, x_2, -h(x_1, x_2, \tilde{t})\big), \tag{45a}$$

we obtain the two tangent vectors

$$\boldsymbol{T}_{b,1} \equiv \frac{\partial \boldsymbol{r}_b}{\partial x_1}, \qquad \boldsymbol{T}_{b,2} \equiv \frac{\partial \boldsymbol{r}_b}{\partial x_2}, \qquad b \in \{\eta, h\}, \tag{45b}$$

which constitute a natural local basis of the tangent plane at each point of the boundary surface. As concerns the lateral rigid-wall boundary, being a fixed vertical surface, it has $\boldsymbol{n}_{\mathrm{w}}(\boldsymbol{x}) = (n_{\mathrm{w},1}, n_{\mathrm{w},2}, 0)(\boldsymbol{x})$, and respective unit tangent vectors

$$\boldsymbol{T}_{\mathrm{w},1} \equiv (0,0,1), \qquad \boldsymbol{T}_{\mathrm{w},2} \equiv (n_{\mathrm{w},2}, -n_{\mathrm{w},1}, 0). \tag{45c}$$

Therefore, the tangent component of the virtual displacements, $\delta_L \boldsymbol{X}_{b,\|}$, may be generally written as

$$\delta_L \boldsymbol{X}_{b,\|} = \delta B_{b,1}\, \boldsymbol{T}_{b,1} + \delta B_{b,2}\, \boldsymbol{T}_{b,2}, \qquad b \in \{\eta, h, \mathrm{w}\}, \tag{46}$$

where $\delta B_{b,1}$ and $\delta B_{b,2}$ are appropriate scalar quantities. Adopting this decomposition for $\delta_L \boldsymbol{X}_b$, leads to the consideration of independent variations of the scalars $\delta B_{b,\{\perp,1,2\}}$ instead of the component variations $(\delta_L X_{b,1}, \delta_L X_{b,2}, \delta_L Z_b)$.

### 5.3. *Free-surface conditions*

Introducing the representations (43) and (46) into the free-surface variational Eq. (40), and considering first tangential variations to the boundary ($\delta_L \boldsymbol{X}_{\eta,\|} =$ arbitrary, $\delta_L \boldsymbol{X}_{\eta,\perp} = 0$), we obtain ([14])

$$\begin{aligned}\int_{\mathrm{T}}\int_{D} \left[\rho k \left(\frac{\partial \boldsymbol{N}_\eta}{\partial t} + \nabla(\boldsymbol{u}\boldsymbol{N}_\eta)\right) - \left(\frac{\partial \eta}{\partial t} - \boldsymbol{u}\boldsymbol{N}_\eta\right)(k\nabla\rho + \rho \boldsymbol{A}\nabla\boldsymbol{a})\right]_{z=\eta} \delta_L \boldsymbol{X}_{\eta,\|}\, d\boldsymbol{x}\, dt \\ - \underbrace{\int_{\mathrm{T}}\int_{D} \left[\rho k \left(\frac{\partial \eta}{\partial t} - \boldsymbol{u}\boldsymbol{N}_\eta\right)\right]_{z=\eta} \nabla_2 \cdot (\delta B_{\eta,1}, \delta B_{\eta,2})\, d\boldsymbol{x}\, dt}_{I_\eta} = 0.\end{aligned} \tag{47}$$

Invoking the 2D divergence theorem for the integral $I_\eta$ of Eq. (47), and neglecting subsequent terms on the line boundary of the free surface (boundary of co-dimension 2), we find

---

([13]) On the free surface and the seabed, it is convenient to express the normal direction with the vector $\boldsymbol{N}_b / \|\boldsymbol{N}_b\|^2$, instead of the unit vector $\boldsymbol{n}_b = \boldsymbol{N}_b / \|\boldsymbol{N}_b\|$. This is done to simplify some expressions in the subsequent algebraic manipulations. Its effect is an indifferent rescaling of the scalar $\delta B_{b,\perp}$.

([14]) Note that $\delta_L \boldsymbol{X}_{\eta,\|} \cdot \boldsymbol{N}_\eta = 0$.

$$I_\eta = -\int_D \nabla_2\left(\left[\rho\, k\left(\frac{\partial\eta}{\partial t} - \boldsymbol{u}\,\boldsymbol{N}_\eta\right)\right]_{z=\eta}\right)(\delta B_{\eta,1}, \delta B_{\eta,2})\, d\boldsymbol{x}.$$

Notice that the above integrand involves a term where the evaluation on $z = \eta$ precedes the differentiation with $\nabla_2(\bullet)$. This originates from the use of Eqs. (35), (36) and (38), for the *a priori* evaluated field variations on the boundary, and it is in contrast to the situation observed in the first integral of Eq. (47), where similar terms are first differentiated and afterwards evaluated on the free surface. The following lemma will facilitate the calculation of terms $\nabla_2(\bullet)$, appearing in the integrand of $I_\eta$ and in the integrand of a similar integral associated with the seabed, below.

**Lemma 1.** For any sufficiently smooth function $f = f(\boldsymbol{x}, z, t)$ of the flow, it holds that

$$\nabla_2\left([f]_{z=\{\eta,-h\}}\right)(\delta B_{\{\eta,h\},1}, \delta B_{\{\eta,h\},2}) = \\ = [\nabla f]_{z=\{\eta,-h\}}(\delta B_{\{\eta,h\},1}\,\boldsymbol{T}_{\{\eta,h\},1} + \delta B_{\{\eta,h\},2}\,\boldsymbol{T}_{\{\eta,h\},2}).$$

The proof of the lemma is given in Appendix B.

Given the above Lemma 1, $I_\eta$ becomes

$$I_\eta = -\int_D \left\{\nabla\left[\rho\, k\left(\frac{\partial\eta}{\partial t} - \boldsymbol{u}\,\boldsymbol{N}_\eta\right)\right]\right\}_{z=\eta}(\delta B_{\eta,1}\,\boldsymbol{T}_{\eta,1} + \delta B_{\eta,2}\,\boldsymbol{T}_{\eta,2})\, d\boldsymbol{x}.$$

Substituting the latter into Eq. (47), and using the following relation ([15])

$$\frac{\partial \boldsymbol{N}_\eta}{\partial t} + \nabla(\boldsymbol{u}\,\boldsymbol{N}_\eta) + \nabla\left(\frac{\partial\eta}{\partial t} - \boldsymbol{u}\,\boldsymbol{N}_\eta\right) = \frac{\partial \boldsymbol{N}_\eta}{\partial t} + \nabla\left(\frac{\partial\eta}{\partial t}\right) = 0,$$

we obtain the final form of the free-surface variational equation for tangential virtual displacements, reading as

$$\int_T\int_D \left[\rho\left(\frac{\partial\eta}{\partial t} - \boldsymbol{u}\,\boldsymbol{N}_\eta\right)(-\nabla k + \boldsymbol{A}\nabla\boldsymbol{a})\right]_{z=\eta}(\delta B_{\eta,1}\,\boldsymbol{T}_{\eta,1} + \delta B_{\eta,2}\,\boldsymbol{T}_{\eta,2})\, d\boldsymbol{x}\, dt = 0.$$

Therefore, since $\rho \neq 0$, $\boldsymbol{u} = -\nabla k + \boldsymbol{A}\nabla\boldsymbol{a}$, and the variations $\delta B_{\eta,1}$ and $\delta B_{\eta,2}$ are independent and arbitrary, we obtain the Euler-Lagrange equations

$$\left(\frac{\partial\eta}{\partial t} - \boldsymbol{u}\,\boldsymbol{N}_\eta\right)\boldsymbol{u}\,\boldsymbol{T}_{\eta,i} = 0, \qquad i = 1,2, \qquad z = \eta(\boldsymbol{x}, t). \tag{48}$$

Eqs. (48) provide us with two (non-exclusive) possibilities: either $\partial\eta/\partial t - \boldsymbol{u}\,\boldsymbol{N}_\eta = 0$ or $u_{\|} =$ magnitude of the tangential velocity $= 0$. Since, on the free-surface, the tangent velocity cannot generally be zero, we conclude that

([15]) Recall Eq. (2a) for $\boldsymbol{N}_\eta$.

$$\delta_L \boldsymbol{X}_{\eta,\parallel} : \qquad \frac{\partial \eta}{\partial t} - \boldsymbol{u}\,\boldsymbol{N}_\eta = 0\,, \qquad\qquad z = \eta\,, \tag{49}$$

which constitutes the *free-surface kinematic condition*, coinciding with Eq. (6).

Now, we return to the free-surface variational Eq. (40), considering normal variations to the boundary ($\delta_L \boldsymbol{X}_{\eta,\perp} =$ arbitrary, $\delta_L \boldsymbol{X}_{\eta,\parallel} = 0$). Given Eq. (44) and the kinematic condition Eq. (49), which also implies that $\partial \boldsymbol{N}_\eta / \partial t + \nabla(\boldsymbol{u}\,\boldsymbol{N}_\eta) = 0$, Eq. (40) becomes

$$\int_{\mathrm{T}}\int_{D} \left[ \rho\left(\frac{\boldsymbol{u}^2}{2} - E - P\right) - k\left(\frac{\partial \rho}{\partial t} + \nabla\cdot(\rho\,\boldsymbol{u})\right) - \rho\,\boldsymbol{A}\frac{D\boldsymbol{a}}{Dt} - \overline{p}\right]_{z=\eta} \delta B_{\eta,\perp}\, d\boldsymbol{x}\, dt$$
$$- \underbrace{\int_{\mathrm{T}}\int_{D} [\rho\, k]_{z=\eta} \frac{\partial}{\partial t}(\delta B_{\eta,\perp})\, d\boldsymbol{x}\, dt}_{J_{\eta,1}} - \underbrace{\int_{\mathrm{T}}\int_{D} \left[\rho\, k\,(u_1,u_2)\right]_{z=\eta} \nabla_2(\delta B_{\eta,\perp})\, d\boldsymbol{x}\, dt}_{J_{\eta,2}} = 0.$$

Using the isochronality condition for the integral $J_{\eta,1}$, and the 2D divergence theorem for the integral $J_{\eta,2}$, we obtain

$$\begin{aligned} &\int_{\mathrm{T}}\int_{D} \left[ \rho\left(\frac{\boldsymbol{u}^2}{2} - E - P\right) - k\left(\frac{\partial \rho}{\partial t} + \nabla\cdot(\rho\,\boldsymbol{u})\right) - \rho\,\boldsymbol{A}\frac{D\boldsymbol{a}}{Dt} - \overline{p}\right]_{z=\eta} \delta B_{\eta,\perp}\, d\boldsymbol{x}\, dt \\ &\quad + \int_{\mathrm{T}}\int_{D} \left\{ \frac{\partial}{\partial t}\Big([\rho\, k]_{z=\eta}\Big) + \nabla_2\cdot\Big(\left[\rho\, k\,(u_1,u_2)\right]_{z=\eta}\Big)\right\} \delta B_{\eta,\perp}\, d\boldsymbol{x}\, dt = 0\,, \end{aligned} \tag{50}$$

where, again, terms on the free surface's line boundary are neglected. The following lemma permits us to reformulate the integrand of the second integral in Eq. (50), and simplify the calculations.

**Lemma 2.** Given the free-surface kinematic condition Eq. (49), any sufficiently smooth function $f = f(\boldsymbol{x}, z, t)$ of the flow satisfies the relation

$$\frac{\partial}{\partial t}\Big([f]_{z=\eta}\Big) + \nabla_2\cdot\Big([f\cdot(u_1,u_2)]_{z=\eta}\Big) = \left[\frac{\partial f}{\partial t} + \nabla\cdot(f\,\boldsymbol{u})\right]_{z=\eta}.$$

The proof of the lemma is given in Appendix B.

Applying Lemma 2 for $f = \rho\, k$, and after simple calculations, Eq. (50) is finally shaped as

$$\int_{\mathrm{T}}\int_{D} \left\{ \rho\frac{Dk}{Dt} - \rho\,\boldsymbol{A}\frac{D\boldsymbol{a}}{Dt} + \rho\left(\frac{\boldsymbol{u}^2}{2} - E - P\right) - \overline{p}\right\}_{z=\eta} \delta B_{\eta,\perp}\, d\boldsymbol{x}\, dt = 0\,.$$

Thus, for arbitrary variations $\delta B_{\eta,\perp}$, we obtain the *free-surface dynamic condition*

$$\delta_L \boldsymbol{X}_{\eta,\perp} : \qquad -\frac{Dk}{Dt} + \boldsymbol{A}\frac{D\boldsymbol{a}}{Dt} - \frac{\boldsymbol{u}^2}{2} + E + P = -\frac{\overline{p}}{\rho}\,, \qquad\qquad z = \eta\,. \tag{51}$$

Recalling the pressure representation Eq. (33a), the above condition is rewritten in the form $p = \overline{p}$, coinciding with Eq. (7).

**Remark 4:** If we take into account the velocity representation (32), then the free-surface dynamic condition, Eq. (51), is written as

$$-\frac{\partial k}{\partial t} + \boldsymbol{A}\frac{\partial \boldsymbol{a}}{\partial t} + \frac{\boldsymbol{u}^2(k, A_i, a_i)}{2} + E + P = -\frac{\overline{p}}{\rho}, \qquad z = \eta, \tag{52}$$

which, recalling the pressure representation Eq. (33b), becomes again $p = \overline{p}$. The left-hand side of Eq. (52) is closely related to the Lagrangian density proposed by Clebsch [29], for the variational formulation of Euler equations, using Clebsch potentials; see, also, [55]. What differs, obviously, is the presence of additional terms, which are owed to the additional features of compressibility, conservative body forces, and applied pressure on the free surface.

**Remark 5:** The free-surface part of the problem is also dealt with by Timokha [32] and Berdichevsky [26] (Sec. 9.3), but for incompressible fluids. (a) Timokha uses the unconstrained Clebsch-Bateman-Luke principle (not Hamilton's principle, as herein), for the case of wave sloshing, where the Lagrangian density is a modification of Eq. (52), involving a single label (original Clebsch representation). (b) Berdichevsky begins with Hamilton's principle, but *a priori imposes the free-surface kinematic condition* as essential. Also, he does not explicitly introduce the free-surface elevation $\eta$. To arrive at the dynamic condition on the free surface, he identifies the connection between the variations of the velocity and of the free-surface boundary, via Eq. (19c). His result is similar to Eq. (51), but without the term $\boldsymbol{A}\, D\boldsymbol{a}/Dt$, which is added artificially (on the basis of the identity conservation) in order to derive Eq. (52).

### 5.4. *Seabed and lateral rigid-wall conditions*

In the seabed variational Eq. (41), Eq. (43) reduces to $\delta_L \boldsymbol{X}_h = \delta_L \boldsymbol{X}_{h,\parallel}$, because of Eq. (21). That is, on the seabed we have to consider only tangential variations, for which Eq. (41) reads

$$\int_{\mathrm{T}}\int_{D}\left\{\left[\rho\, k\left(\frac{\partial \boldsymbol{N}_h}{\partial t} + \nabla(\boldsymbol{u}\,\boldsymbol{N}_h)\right) - \left(\frac{\partial h}{\partial t} - \boldsymbol{u}\,\boldsymbol{N}_h\right)(k\nabla\rho + \rho\,\boldsymbol{A}\nabla\boldsymbol{a})\right]_{z=-h}\delta_L \boldsymbol{X}_{h,\parallel} \right.$$
$$\left. - \left[\rho\, k\left(\frac{\partial h}{\partial t} - \boldsymbol{u}\,\boldsymbol{N}_h\right)\right]_{z=-h}\nabla_2\cdot(\delta B_{h,1}, \delta B_{h,2})\right\} d\boldsymbol{x}\, dt = 0,$$

with $\delta_L \boldsymbol{X}_{h,\parallel} = \delta B_{h,1}\boldsymbol{T}_{h,1} + \delta B_{h,2}\boldsymbol{T}_{h,2}$; see Eqs. (45a,b) – (46). This variational equation is of exactly the same structure as Eq. (47) above, having the index $h$ in the place of $\eta$, and with evaluation of the terms on $z = -h$ in the place of $z = \eta$. Thus, we treat it in a similar way, concluding to the final form

$$\int_{\mathrm{T}}\int_{D}\left[\rho\left(\frac{\partial h}{\partial t} - \boldsymbol{u}\,\boldsymbol{N}_h\right)(-\nabla k + \boldsymbol{A}\nabla\boldsymbol{a})\right]_{z=-h}(\delta B_{h,1}\boldsymbol{T}_{h,1} + \delta B_{h,2}\boldsymbol{T}_{h,2})\, d\boldsymbol{x}\, dt = 0.$$

As before, since $\boldsymbol{u} = -\nabla k + \boldsymbol{A}\nabla\boldsymbol{a}$, and the variations $\delta B_{h,1}$ and $\delta B_{h,2}$ are independent and arbitrary, we obtain the Euler-Lagrange equations

$$\left( \frac{\partial h}{\partial t} - \boldsymbol{u}\, \boldsymbol{N}_h \right) \boldsymbol{u}\, \boldsymbol{T}_{h,i} = 0, \qquad i = 1,2, \qquad z = -h(\boldsymbol{x},t). \tag{53}$$

Thus, thinking as in the case of the free surface, we conclude to the *moving-seabed kinematic condition*

$$\delta_L \boldsymbol{X}_{h,\parallel} : \quad \frac{\partial h}{\partial t} - \boldsymbol{u}\, \boldsymbol{N}_h = 0, \qquad z = -h. \tag{54}$$

Obviously, this condition coincides with Eq. (8).

Last, on the lateral rigid wall, owing to Eq. (22), the variations are again only tangential, that is, $\delta_L \boldsymbol{X}_{\mathrm{w}} = \delta_L \boldsymbol{X}_{\mathrm{w},\parallel}$, given by Eq. (45c) – (46). Accordingly, after simple algebraic calculations, the variational Eq. (42) becomes

$$-\int_{\mathrm{T}} \int_{\partial D_{\mathrm{w}}} \int_{-h(\boldsymbol{x},t)}^{\eta(\boldsymbol{x},t)} \rho\,(\boldsymbol{u}\, \boldsymbol{n}_{\mathrm{w}})(-\nabla k + \boldsymbol{A} \nabla \boldsymbol{a})\, \delta_L \boldsymbol{X}_{\mathrm{w},\parallel}\, dz\, dl\, dt$$

$$-\int_{\mathrm{T}} \int_{\partial D_{\mathrm{w}}} \int_{-h(\boldsymbol{x},t)}^{\eta(\boldsymbol{x},t)} \nabla \cdot \{\rho\, k\, (\boldsymbol{u}\, \boldsymbol{n}_{\mathrm{w}})\, \delta_L \boldsymbol{X}_{\mathrm{w},\parallel}\}\, dz\, dl\, dt = 0.$$

The second integral, though, is just the divergence of a vector field over the surface $\partial V_{\mathrm{w}}$. Thus, it is omitted, integrating out to the line boundary of the lateral rigid wall. As a result, we get the following Euler-Lagrange equations

$$(\boldsymbol{u}\, \boldsymbol{n}_{\mathrm{w}})\, \boldsymbol{u}\, \boldsymbol{T}_{\mathrm{w},i} = 0, \qquad i = 1,2, \qquad (\boldsymbol{x},z) \in \partial V_{\mathrm{w}}, \tag{55}$$

which lead to the *impermeability condition* Eq. (9),

$$\delta_L \boldsymbol{X}_{\mathrm{w},\parallel} : \quad \boldsymbol{u}\, \boldsymbol{n}_{\mathrm{w}} = 0, \qquad (\boldsymbol{x},z) \in \partial V_{\mathrm{w}}, \tag{56}$$

using the same reasoning as previously. With the latter condition, we complete the variational derivation of all boundary conditions that are involved in the differential formulation of the problem (Sec. 2.3), using Hamilton's Principle.

**Remark 6:** Interestingly enough, we see that the variational derivation of kinematic boundary conditions for all the boundaries leads to a dual possibility:

> *Either* the tangential velocity on the considered boundary is identically zero,
> *or* the usual kinematic condition (the same as in irrotational flow) holds true.

See Eqs. (48), (53) and (55). Without being able to discuss in depth the above duality, we mention that the zero tangential velocity is a boundary condition appropriate for the Navier-Stokes equations. This makes plausible the conjecture that the zero-viscosity limit of Navier-Stokes equations can match with solutions of Euler equations exhibiting zero tangential components.

## 6. Discussion and conclusions

In this work, the Herivel-Lin variational approach for rotational flows, based on Hamilton's Principle, has been extended to the case of an ideal barotropic fluid with a free surface and a moving seabed. The variational derivation of boundary conditions for such problems seems to be lacking from the relevant literature. A possible explanation for this situation might be the inadequacy of Lin's constraint to resolve the variational controversy of the Eulerian description on the boundary, although it does so for the bulk fluid motion. To address this issue, additional differential-variational constraints are introduced for the Eulerian fields' variations on the boundary, leading to a reformulation of the boundary variational equation, in terms of the virtual displacements of the boundary fluid parcels. In this way, the variational controversy is dealt with on the boundaries, as well, and correct boundary conditions are derived.

This, somewhat unconventional, variational procedure succeeds in two main points. First, besides the derivation of the full equations for the bulk fluid motion, convenient representations of the velocity and pressure fields are obtained in terms of Clebsch-Lin potentials. This is possible because of the ability to consider independent variations of the Eulerian fields in the interior of the fluid domain, owing to the inclusion of the mass and identity constraints into the action functional. Second, re-expressing the boundary variational equation in terms of the boundary fluid parcels' virtual displacements, leads to a successful derivation of both kinematic and dynamic boundary conditions. To our knowledge, this has never been done before, for such flows, using Hamilton's Principle.

The present variational formulation, however complicated might be, offers valuable information concerning the variational treatment of rotational free-surface flows. Though, certain concepts that are involved require further clarification. For example, i) the two-level constraints (different in the fluid volume and on the fluid boundary), introduced herein for the first time, call for a better understanding and theoretical assessment; ii) the gauge freedom of the Clebsch-Lin potentials and the ability of the latter to represent generic rotational flows need to be clarified. It is the authors' intention to address those issues in the near future. Nevertheless, we believe that the outcome of this work is a decisive step towards a subsequent formulation of an unconstrained variational principle, which will incorporate all the essential features of rotational free-surface flows.

## Appendix A. Detailed calculation of the action functional's partial Gateaux derivatives

In this Appendix, we derive the partial Gateaux derivatives (26c,d,e), of the action functional Eq. (16), Sec. 4.1, with respect to the fields $\boldsymbol{a}(\boldsymbol{x},z,t)$, $\rho(\boldsymbol{x},z,t)$ and $\boldsymbol{u}(\boldsymbol{x},z,t)$.

***Variation of the parcel labels*:** Considering the variation $\delta_{\boldsymbol{a}}(\bullet)$ of the parcel labels, for the action functional Eq. (16), we have that

$$\begin{aligned}\delta_{\boldsymbol{a}}\tilde{\mathscr{S}} &= -\int_{\mathrm{T}}\int_{D}\int_{-h(\boldsymbol{x},t)}^{\eta(\boldsymbol{x},t)} \rho A\frac{D(\delta\boldsymbol{a})}{Dt}\,dz\,d\boldsymbol{x}\,dt = \\ &= \int_{\mathrm{T}}\int_{D}\int_{-h(\boldsymbol{x},t)}^{\eta(\boldsymbol{x},t)} \frac{D(\rho A)}{Dt}\delta\boldsymbol{a}\,dz\,d\boldsymbol{x}\,dt \\ &\quad - \underbrace{\int_{\mathrm{T}}\int_{D}\int_{-h(\boldsymbol{x},t)}^{\eta(\boldsymbol{x},t)} \frac{D}{Dt}(\rho A\,\delta\boldsymbol{a})\,dz\,d\boldsymbol{x}\,dt}_{I}.\end{aligned} \tag{A1}$$

The second integral of the rightmost-hand side above equation, $I$, can be further analyzed as

$$\begin{aligned}I &= \int_{\mathrm{T}}\int_{D}\int_{-h(\boldsymbol{x},t)}^{\eta(\boldsymbol{x},t)} \left(\frac{\partial}{\partial t}(\rho A\,\delta\boldsymbol{a}) + \boldsymbol{u}\nabla(\rho A\,\delta\boldsymbol{a})\right)dz\,d\boldsymbol{x}\,dt = \\ &= \int_{\mathrm{T}}\int_{D}\underbrace{\int_{-h(\boldsymbol{x},t)}^{\eta(\boldsymbol{x},t)} \frac{\partial}{\partial t}(\rho A\,\delta\boldsymbol{a})\,dz}_{I_1}\,d\boldsymbol{x}\,dt + \int_{\mathrm{T}}\int_{D}\underbrace{\int_{-h(\boldsymbol{x},t)}^{\eta(\boldsymbol{x},t)} \boldsymbol{u}\nabla(\rho A\,\delta\boldsymbol{a})\,dz}_{I_2}\,d\boldsymbol{x}\,dt.\end{aligned}$$

Then, utilizing Leibnitz integral rule for differentiation under the integral sign, and also using the relation $\boldsymbol{u}\nabla(\rho A\,\delta\boldsymbol{a}) = \nabla\cdot\big((\rho A\,\delta\boldsymbol{a})\boldsymbol{u}\big) - (\nabla\cdot\boldsymbol{u})\,\rho A\,\delta\boldsymbol{a}$ for $I_2$, we obtain

$$I_1 = \frac{\partial}{\partial t}\int_{-h(\boldsymbol{x},t)}^{\eta(\boldsymbol{x},t)} \rho A\,\delta\boldsymbol{a}\,dz - \left[\rho A\,\delta\boldsymbol{a}\right]_{z=\eta}\frac{\partial\eta}{\partial t} - \left[\rho A\,\delta\boldsymbol{a}\right]_{z=-h}\frac{\partial h}{\partial t}$$

and

$$\begin{aligned}I_2 &= \int_{-h(\boldsymbol{x},t)}^{\eta(\boldsymbol{x},t)} \nabla\cdot[(\rho A\,\delta\boldsymbol{a})\boldsymbol{u}]\,dz - \int_{-h(\boldsymbol{x},t)}^{\eta(\boldsymbol{x},t)} (\nabla\cdot\boldsymbol{u})\,\rho A\,\delta\boldsymbol{a}\,dz = \\ &= \sum_{i=1}^{2}\int_{-h(\boldsymbol{x},t)}^{\eta(\boldsymbol{x},t)} \frac{\partial}{\partial x_i}[(\rho A\,\delta\boldsymbol{a})u_i]\,dz + \int_{-h(\boldsymbol{x},t)}^{\eta(\boldsymbol{x},t)} \frac{\partial}{\partial z}[(\rho A\,\delta\boldsymbol{a})u_3]\,dz \\ &\quad - \int_{-h(\boldsymbol{x},t)}^{\eta(\boldsymbol{x},t)} (\nabla\cdot\boldsymbol{u})\,\rho A\,\delta\boldsymbol{a}\,dz =\end{aligned}$$

$$= \sum_{i=1}^{2} \left( \frac{\partial}{\partial x_i} \int_{-h(\boldsymbol{x},t)}^{\eta(\boldsymbol{x},t)} (\rho A\, \delta \boldsymbol{a})\, u_i \, dz \right.$$

$$\left. - \big[ (\rho A\, \delta \boldsymbol{a})\, u_i \big]_{z=\eta} \frac{\partial \eta}{\partial x_i} - \big[ (\rho A\, \delta \boldsymbol{a})\, u_i \big]_{z=-h} \frac{\partial h}{\partial x_i} \right)$$

$$+ \big[ (\rho A\, \delta \boldsymbol{a})\, u_3 \big]_{z=\eta} - \big[ (\rho A\, \delta \boldsymbol{a})\, u_3 \big]_{z=-h} - \int_{-h(\boldsymbol{x},t)}^{\eta(\boldsymbol{x},t)} (\nabla \cdot \boldsymbol{u})\, \rho A\, \delta \boldsymbol{a}\, dz .$$

Combining the above results, and rearranging various terms, the integral $I$ finally becomes

$$I = \int_{\mathrm{T}} \int_{D} \left\{ \frac{\partial}{\partial t} \int_{-h(\boldsymbol{x},t)}^{\eta(\boldsymbol{x},t)} \rho A\, \delta \boldsymbol{a}\, dz + \sum_{i=1}^{2} \frac{\partial}{\partial x_i} \int_{-h(\boldsymbol{x},t)}^{\eta(\boldsymbol{x},t)} (\rho A\, \delta \boldsymbol{a})\, u_i \, dz \right.$$

$$- \int_{-h(\boldsymbol{x},t)}^{\eta(\boldsymbol{x},t)} (\nabla \cdot \boldsymbol{u})\, \rho A\, \delta \boldsymbol{a}\, dz - \left( \frac{\partial \eta}{\partial t} - [\boldsymbol{u}]_{z=\eta} \boldsymbol{N}_{\eta} \right) \big[ \rho A\, \delta \boldsymbol{a} \big]_{z=\eta}$$

$$\left. - \left( \frac{\partial h}{\partial t} - [\boldsymbol{u}]_{z=-h} \boldsymbol{N}_{h} \right) \big[ \rho A\, \delta \boldsymbol{a} \big]_{z=-h} \right\} d\boldsymbol{x}\, dt, \qquad \text{(A2)}$$

recalling that $\boldsymbol{N}_{\eta}$ and $\boldsymbol{N}_{h}$ are outward normal vectors on the free surface and the seabed, respectively, given by Eqs. (2a) and (2b).

However, the first integral of the above equation integrates out to the boundaries of the time domain $\mathrm{T}$ and, thus, vanishes, according to the isochronality condition. That is,

$$\int_{\mathrm{T}} \int_{D} \frac{\partial}{\partial t} \int_{-h(\boldsymbol{x},t)}^{\eta(\boldsymbol{x},t)} \rho A\, \delta \boldsymbol{a}\, dz\, d\boldsymbol{x}\, dt = 0 . \qquad \text{(A3)}$$

As for the next two integrals, we may write

$$I_{\partial V_{\mathrm{lat}}} \equiv \int_{\mathrm{T}} \int_{D} \sum_{i=1}^{2} \frac{\partial}{\partial x_i} \int_{-h(\boldsymbol{x},t)}^{\eta(\boldsymbol{x},t)} (\rho A\, \delta \boldsymbol{a})\, u_i \, dz\, d\boldsymbol{x}\, dt =$$
$$= \int_{\mathrm{T}} \int_{D} \nabla_2 \cdot \boldsymbol{G}(\boldsymbol{x},t)\, d\boldsymbol{x}\, dt, \qquad \text{(A4)}$$

$\nabla_2(\bullet) \equiv \big( \partial(\bullet)/\partial x_1, \partial(\bullet)/\partial x_2 \big)$ being the 2D gradient, where

$$\boldsymbol{G}(\boldsymbol{x},t) = \big( G_1(\boldsymbol{x},t), G_2(\boldsymbol{x},t) \big), \qquad G_i(\boldsymbol{x},t) = \int_{-h(\boldsymbol{x},t)}^{\eta(\boldsymbol{x},t)} (\rho A\, \delta \boldsymbol{a})\, u_i \, dz .$$

Then, invoking the divergence theorem in two dimensions, we obtain

$$I_{\partial V_{\text{lat}}} = \int_{\mathrm{T}}\int_{D} \nabla_2 \cdot \boldsymbol{G}(\boldsymbol{x},t)\, d\boldsymbol{x}\, dt = \int_{\mathrm{T}}\oint_{\partial D} \boldsymbol{G}(\boldsymbol{x},t)\, \boldsymbol{n}_{\partial D}\, dl\, dt,$$

where $\boldsymbol{n}_{\partial D}(\boldsymbol{x}) = (n_{\partial D,1}, n_{\partial D,2})(\boldsymbol{x})$ is the outward unit normal vector on the boundary $\partial D$ of the horizontal domain. Introducing, next, the explicit form of $\boldsymbol{G}(\boldsymbol{x},t)$ into the above, and rearranging terms, $I_{\partial V_{\text{lat}}}$ is written as

$$I_{\partial V_{\text{lat}}} = \int_{\mathrm{T}}\oint_{\partial D} \left\{ \int_{-h(\boldsymbol{x},t)}^{\eta(\boldsymbol{x},t)} (\rho A\, \delta\boldsymbol{a})\, u_1\, n_{\partial D,1}\, dz + \int_{-h(\boldsymbol{x},t)}^{\eta(\boldsymbol{x},t)} (\rho A\, \delta\boldsymbol{a})\, u_2\, n_{\partial D,2}\, dz \right\} dl\, dt =$$

$$= \int_{\mathrm{T}}\oint_{\partial D} \int_{-h(\boldsymbol{x},t)}^{\eta(\boldsymbol{x},t)} (\rho A\, \delta\boldsymbol{a}) \left\{ u_1\, n_{\partial D,1} + u_2\, n_{\partial D,2} \right\} dz\, dl\, dt =$$

$$= \int_{\mathrm{T}}\oint_{\partial D} \int_{-h(\boldsymbol{x},t)}^{\eta(\boldsymbol{x},t)} (\rho A\, \delta\boldsymbol{a})\, \boldsymbol{u}\, \boldsymbol{n}_{\text{lat}}\, dz\, dl\, dt, \tag{A5}$$

where $\boldsymbol{n}_{\text{lat}}(\boldsymbol{x}) = (n_{\partial D,1}, n_{\partial D,2}, 0)(\boldsymbol{x})$ is the outward unit normal vector on the (vertical) lateral surface $\partial V_{\text{lat}}$. Using, therefore, Eqs. (A2) – (A5) into Eq. (A1), and given the fact that $\delta\boldsymbol{a}$ vanish everywhere on $\partial V_{\text{lat}}$, except for the rigid-wall part $\partial V_{\text{w}}$ (see Sec. 2.3 and 3.4), we finally obtain Eq. (26c).

***Variation of the fluid density***: For the variation $\delta_\rho(\bullet)$ of the action functional Eq. (16), we initially calculate

$$\delta_\rho \tilde{\mathscr{S}} = \int_{\mathrm{T}}\int_{D}\int_{-h(\boldsymbol{x},t)}^{\eta(\boldsymbol{x},t)} \left\{ \left( \frac{1}{2}\boldsymbol{u}^2 - E - P \right) \delta\rho - \rho \frac{\partial E}{\partial \rho} \delta\rho - k\left( \frac{\partial(\delta\rho)}{\partial t} + \boldsymbol{u}\cdot\nabla(\delta\rho) + \nabla\cdot\boldsymbol{u}\,\delta\rho \right) - A\frac{D\boldsymbol{a}}{Dt}\delta\rho \right\} dz\, d\boldsymbol{x}\, dt =$$

$$= \int_{\mathrm{T}}\int_{D}\int_{-h(\boldsymbol{x},t)}^{\eta(\boldsymbol{x},t)} \left( \frac{1}{2}\boldsymbol{u}^2 - E - P - \rho\frac{\partial E}{\partial \rho} - k\,\nabla\cdot\boldsymbol{u} - A\frac{D\boldsymbol{a}}{Dt} \right) \delta\rho\, dz\, d\boldsymbol{x}\, dt$$

$$- \underbrace{\int_{\mathrm{T}}\int_{D}\int_{-h(\boldsymbol{x},t)}^{\eta(\boldsymbol{x},t)} k\,\frac{\partial(\delta\rho)}{\partial t}\, dz\, d\boldsymbol{x}\, dt}_{J_1} - \underbrace{\int_{\mathrm{T}}\int_{D}\int_{-h(\boldsymbol{x},t)}^{\eta(\boldsymbol{x},t)} k\,\boldsymbol{u}\cdot\nabla(\delta\rho)\, dz\, d\boldsymbol{x}\, dt}_{J_2}. \tag{A6}$$

As concerns the above integrals $J_1$ and $J_2$, using the Leibnitz integral rule and the isochronality condition, we get

$$J_1 = \int\limits_{-h(\boldsymbol{x},t)}^{\eta(\boldsymbol{x},t)} \frac{\partial}{\partial t}(k\,\delta\rho)\,dz - \int\limits_{-h(\boldsymbol{x},t)}^{\eta(\boldsymbol{x},t)} \frac{\partial k}{\partial t}\delta\rho\,dz =$$

$$= \cancel{\frac{\partial}{\partial t}\int\limits_{-h(\boldsymbol{x})}^{\eta(\boldsymbol{x},t)} k\,\delta\rho\,dz} - \left[k\,\delta\rho\right]_{z=\eta}\frac{\partial\eta}{\partial t} - \left[k\,\delta\rho\right]_{z=-h}\frac{\partial h}{\partial t} - \int\limits_{-h(\boldsymbol{x},t)}^{\eta(\boldsymbol{x},t)} \frac{\partial k}{\partial t}\delta\rho\,dz,$$

and

$$J_2 = \int\limits_{-h(\boldsymbol{x},t)}^{\eta(\boldsymbol{x},t)} \nabla\cdot(k\,\boldsymbol{u}\,\delta\rho)\,dz - \int\limits_{-h(\boldsymbol{x},t)}^{\eta(\boldsymbol{x},t)} \nabla\cdot(k\,\boldsymbol{u})\,\delta\rho\,dz =$$

$$= \sum_{i=1}^{2}\int\limits_{-h(\boldsymbol{x},t)}^{\eta(\boldsymbol{x},t)} \frac{\partial}{\partial x_i}(k\,u_i\,\delta\rho)\,dz + \int\limits_{-h(\boldsymbol{x},t)}^{\eta(\boldsymbol{x},t)} \frac{\partial}{\partial z}(k\,u_3\,\delta\rho)\,dz$$

$$- \int\limits_{-h(\boldsymbol{x},t)}^{\eta(\boldsymbol{x},t)} \nabla\cdot(k\,\boldsymbol{u})\,\delta\rho\,dz =$$

$$= \sum_{i=1}^{2}\left(\underline{\frac{\partial}{\partial x_i}\int\limits_{-h(\boldsymbol{x},t)}^{\eta(\boldsymbol{x},t)} k\,u_i\,\delta\rho\,dz} - \left[k\,u_i\,\delta\rho\right]_{z=\eta}\frac{\partial\eta}{\partial x_i} - \left[k\,u_i\,\delta\rho\right]_{z=-h}\frac{\partial h}{\partial x_i}\right)$$

$$+\left[k\,u_3\,\delta\rho\right]_{z=\eta} - \left[k\,u_3\,\delta\rho\right]_{z=-h} - \int\limits_{-h(\boldsymbol{x},t)}^{\eta(\boldsymbol{x},t)} \nabla\cdot(k\,\boldsymbol{u})\,\delta\rho\,dz\,.$$

The underlined terms of the last expression, though, are similar to the terms of $I_{\partial V_{\mathrm{lat}}}$, Eq. (A4), studied above, and are treated in the same way. Thus, performing this analysis and substituting the results for the integrals $J_1$, $J_2$ into Eq. (A6), results, after some simple algebraic manipulations, in Eq. (26d).

***Variation of the velocity field***: Regarding the calculation of $\delta_{\boldsymbol{u}}(\bullet)$, we start with the variations $\delta_{u_i}(\bullet)$, $i=1,2$, of the two horizontal velocity components, obtaining ([16]) ([17])

$$\delta_{u_i}\tilde{\mathscr{S}} = \int\limits_{\mathrm{T}}\int\limits_{D}\int\limits_{-h(\boldsymbol{x},t)}^{\eta(\boldsymbol{x},t)} \left\{\rho\,u_i - k\frac{\partial\rho}{\partial x_i} - \rho\,\boldsymbol{A}\frac{\partial\boldsymbol{a}}{\partial x_i}\right\}\delta u_i\,dz\,d\boldsymbol{x}\,dt$$

$$-\int\limits_{\mathrm{T}}\int\limits_{D}\int\limits_{-h(\boldsymbol{x},t)}^{\eta(\boldsymbol{x},t)} \rho\,k\frac{\partial(\delta u_i)}{\partial x_i}\,dz\,d\boldsymbol{x}\,dt =$$

---

([16]) Repeated indices are not summed. We don't use the summation convention.

([17]) For the last equality, we use the identity $\rho\,k\,\partial(\delta u_i)/\partial x_i = \partial(\rho\,k\,\delta u_i)/\partial x_i - \delta u_i\,\partial(\rho\,k)/\partial x_i$.

$$= \int_{\mathrm{T}}\int_{D}\int_{-h(\boldsymbol{x})}^{\eta(\boldsymbol{x},t)} \left\{ \rho u_i - k\frac{\partial \rho}{\partial x_i} - \rho \boldsymbol{A}\frac{\partial \boldsymbol{a}}{\partial x_i} + \frac{\partial(\rho k)}{\partial x_i} \right\} \delta u_i \, dz\, d\boldsymbol{x}\, dt$$

$$- \int_{\mathrm{T}}\int_{D}\int_{-h(\boldsymbol{x})}^{\eta(\boldsymbol{x},t)} \frac{\partial}{\partial x_i}(\rho k\, \delta u_i)\, dz\, d\boldsymbol{x}\, dt .$$

Hence, exploiting the Leibnitz integral rule, for the treatment of the vertical integral in the last term, the above becomes

$$\delta_{u_i}\tilde{\mathscr{S}} = \int_{\mathrm{T}}\int_{D} \left\{ \int_{-h(\boldsymbol{x},t)}^{\eta(\boldsymbol{x},t)} \left( \rho u_i - k\frac{\partial \rho}{\partial x_i} - \rho \boldsymbol{A}\frac{\partial \boldsymbol{a}}{\partial x_i} + \frac{\partial(\rho k)}{\partial x_i} \right) \delta u_i \, dz \right.$$

$$\left. + \left[\rho k\, \delta u_i\right]_{z=\eta} \frac{\partial \eta}{\partial x_i} + \left[\rho k\, \delta u_i\right]_{z=-h} \frac{\partial h}{\partial x_i} \right\} d\boldsymbol{x}\, dt \qquad \text{(A7)}$$

$$- \underbrace{\int_{\mathrm{T}}\int_{D} \frac{\partial}{\partial x_i} \int_{-h(\boldsymbol{x},t)}^{\eta(\boldsymbol{x},t)} \rho k\, \delta u_i \, dz\, d\boldsymbol{x}\, dt .}_{K_i}$$

Similarly, the calculation of $\delta_{u_3}\tilde{\mathscr{S}}$ yields

$$\delta_{u_3}\tilde{\mathscr{S}} = \int_{\mathrm{T}}\int_{D}\int_{-h(\boldsymbol{x},t)}^{\eta(\boldsymbol{x},t)} \left\{ \rho u_3 - k\frac{\partial \rho}{\partial z} - \rho \boldsymbol{A}\frac{\partial \boldsymbol{a}}{\partial z} \right\} \delta u_3 \, dz\, d\boldsymbol{x}\, dt$$

$$- \int_{\mathrm{T}}\int_{D}\int_{-h(\boldsymbol{x},t)}^{\eta(\boldsymbol{x},t)} \rho k \frac{\partial(\delta u_3)}{\partial z} dz\, d\boldsymbol{x}\, dt =$$

$$= \int_{\mathrm{T}}\int_{D}\int_{-h(\boldsymbol{x},t)}^{\eta(\boldsymbol{x},t)} \left\{ \rho u_3 - k\frac{\partial \rho}{\partial z} - \rho \boldsymbol{A}\frac{\partial \boldsymbol{a}}{\partial z} + \frac{\partial}{\partial z}(\rho k) \right\} \delta u_3 \, dz\, d\boldsymbol{x}\, dt$$

$$- \int_{\mathrm{T}}\int_{D}\int_{-h(\boldsymbol{x},t)}^{\eta(\boldsymbol{x},t)} \frac{\partial}{\partial z}(\rho k\, \delta u_3)\, dz\, d\boldsymbol{x}\, dt =$$

$$= \int_{\mathrm{T}}\int_{D}\int_{-h(\boldsymbol{x},t)}^{\eta(\boldsymbol{x},t)} \left\{ \rho u_3 - k\frac{\partial \rho}{\partial z} - \rho \boldsymbol{A}\frac{\partial \boldsymbol{a}}{\partial z} + \frac{\partial}{\partial z}(\rho k) \right\} \delta u_3 \, dz\, d\boldsymbol{x}\, dt \qquad \text{(A8)}$$

$$- \int_{\mathrm{T}}\int_{D} \left( \left[\rho k\, \delta u_3\right]_{z=\eta} - \left[\rho k\, \delta u_3\right]_{z=-h} \right) d\boldsymbol{x}\, dt .$$

Thus, combining Eqs. (A7) and (A8), and treating the sum of the integrals $K_1$, $K_2$ (last integral of Eq. (A7), for $i = 1,2$ ) in the same manner as the integral $I_{\partial V_{\mathrm{lat}}}$, Eq. (A4), we conclude to Eq. (26e).

**Appendix B. Proofs of Lemmata 1 and 2**

In this Appendix, we provide the proofs of the two lemmata that are used in Sec.5.

***Proof of Lemma* 1:** Let $\tilde{f} \equiv [f]_{z=\eta} = f\left(\boldsymbol{x},\eta(\boldsymbol{x},t),t\right)$. Then, using the chain rule,

$$\nabla_2 \tilde{f} = \left( \frac{\partial_e \tilde{f}}{\partial x_1} + \frac{\partial \tilde{f}}{\partial \eta}\frac{\partial \eta}{\partial x_1}, \frac{\partial_e \tilde{f}}{\partial x_2} + \frac{\partial \tilde{f}}{\partial \eta}\frac{\partial \eta}{\partial x_2} \right), \tag{B1}$$

where $\partial_e(\bullet)/\partial x_i$ differs from $\partial(\bullet)/\partial x_i$ in the sense that it acts only on the explicit dependence of $\tilde{f}$ on $x_i$; $\tilde{f}$ depends on $t$, $x_1$ and $x_2$ both explicitly and implicitly via $\eta(\boldsymbol{x},t)$. Also, it may be easily checked that

$$\frac{\partial_e \tilde{f}}{\partial x_i} = \left[\frac{\partial f(\boldsymbol{x},z,t)}{\partial x_i}\right]_{z=\eta}, \quad i=1,2, \qquad \text{and} \qquad \frac{\partial \tilde{f}}{\partial \eta} = \left[\frac{\partial f(\boldsymbol{x},z,t)}{\partial z}\right]_{z=\eta}. \tag{B2}$$

Thus, substituting Eqs. (B2) into Eq. (B1), we find

$$\nabla_2 \tilde{f} = \left( \left[\frac{\partial f}{\partial x_1}\right]_{z=\eta} + \left[\frac{\partial f}{\partial z}\right]_{z=\eta}\frac{\partial \eta}{\partial x_1}, \left[\frac{\partial f}{\partial x_2}\right]_{z=\eta} + \left[\frac{\partial f}{\partial z}\right]_{z=\eta}\frac{\partial \eta}{\partial x_2} \right). \tag{B3}$$

Combining the left-hand side of Lemma 1, Sec. 5.3, with the above relation, we obtain

$$\begin{aligned}
&\nabla_2\left([f]_{z=\eta}\right)(\delta B_{\eta,1},\delta B_{\eta,2}) = \\
&= \left( \left[\frac{\partial f}{\partial x_1}\right]_{z=\eta} + \left[\frac{\partial f}{\partial z}\right]_{z=\eta}\frac{\partial \eta}{\partial x_1}\right)\delta B_{\eta,1} + \left( \left[\frac{\partial f}{\partial x_2}\right]_{z=\eta} + \left[\frac{\partial f}{\partial z}\right]_{z=\eta}\frac{\partial \eta}{\partial x_2}\right)\delta B_{\eta,2} = \\
&= \left[\frac{\partial f}{\partial x_1}\right]_{z=\eta}\delta B_{\eta,1} + \left[\frac{\partial f}{\partial x_2}\right]_{z=\eta}\delta B_{\eta,2} + \left[\frac{\partial f}{\partial z}\right]_{z=\eta}\left(\frac{\partial \eta}{\partial x_1}\delta B_{\eta,1} + \frac{\partial \eta}{\partial x_2}\delta B_{\eta,2}\right) = \\
&= [\nabla f]_{z=\eta}\cdot\left(\delta B_{\eta,1},\ \delta B_{\eta,2},\ \frac{\partial \eta}{\partial x_1}\delta B_{\eta,1} + \frac{\partial \eta}{\partial x_2}\delta B_{\eta,2}\right).
\end{aligned} \tag{B4}$$

Additionally, recalling Eqs. (45a,b),

$$\left(\delta B_{\eta,1},\ \delta B_{\eta,2},\ \frac{\partial \eta}{\partial x_1}\delta B_{\eta,1} + \frac{\partial \eta}{\partial x_2}\delta B_{\eta,2}\right) = \delta B_{\eta,1}\boldsymbol{T}_{\eta,1} + \delta B_{\eta,2}\boldsymbol{T}_{\eta,2}. \tag{B5}$$

Using Eq. (B5) into Eq. (B4) concludes the proof of the lemma.

***Proof of Lemma* 2:** Let $\tilde{f} \equiv [f]_{z=\eta} = f\left(\boldsymbol{x},\eta(\boldsymbol{x},t),t\right)$. Then, the above Eqs. (B1) – (B3) hold, along with

$$\begin{aligned}
\frac{\partial \tilde{f}}{\partial t} &= \frac{\partial}{\partial t}\left([f]_{z=\eta}\right) = \frac{\partial_e \tilde{f}}{\partial t} + \frac{\partial \tilde{f}}{\partial \eta}\frac{\partial \eta}{\partial t} = \\
&= \left[\frac{\partial f(\boldsymbol{x},z,t)}{\partial t}\right]_{z=\eta} + \left[\frac{\partial f(\boldsymbol{x},z,t)}{\partial z}\right]_{z=\eta}\frac{\partial \eta}{\partial t},
\end{aligned} \tag{B6}$$

where $\partial_e(\bullet)/\partial t$ has similar meaning as $\partial_e(\bullet)/\partial x_i$, in the proof of Lemma 1. Further,

$$\nabla_2 \cdot \left( [f \cdot (u_1, u_2)]_{z=\eta} \right) = \frac{\partial}{\partial x_1}\left( [f\, u_1]_{z=\eta} \right) + \frac{\partial}{\partial x_2}\left( [f\, u_2]_{z=\eta} \right),$$

which, based on Eqs. (B1) and (B2), is written as

$$\begin{aligned}
&\nabla_2 \cdot \left( [f \cdot (u_1, u_2)]_{z=\eta} \right) = \\
&= \left[ \frac{\partial (f u_1)}{\partial x_1} \right]_{z=\eta} + \left[ \frac{\partial (f u_1)}{\partial z} \right]_{z=\eta} \frac{\partial \eta}{\partial x_1} + \left[ \frac{\partial (f u_2)}{\partial x_2} \right]_{z=\eta} + \left[ \frac{\partial (f u_2)}{\partial z} \right]_{z=\eta} \frac{\partial \eta}{\partial x_2} = \\
&= \left[ \frac{\partial (f u_1)}{\partial x_1} + \frac{\partial (f u_2)}{\partial x_2} \right]_{z=\eta} + \left[ \frac{\partial f}{\partial z} \right]_{z=\eta} \left( [u_1]_{z=\eta} \frac{\partial \eta}{\partial x_1} + [u_2]_{z=\eta} \frac{\partial \eta}{\partial x_2} \right) \\
&\qquad + [f]_{z=\eta} \left( \left[ \frac{\partial u_1}{\partial z} \right]_{z=\eta} \frac{\partial \eta}{\partial x_1} + \left[ \frac{\partial u_2}{\partial z} \right]_{z=\eta} \frac{\partial \eta}{\partial x_2} \right).
\end{aligned} \tag{B7}$$

Accordingly, combining Eqs. (B6) – (B7), the left-hand side of Lemma 2, Sec. 5.3, is equal to

$$\begin{aligned}
&\frac{\partial}{\partial t}\left( [f]_{z=\eta} \right) + \nabla_2 \cdot \left( [f \cdot (u_1, u_2)]_{z=\eta} \right) = \\
&= \left[ \frac{\partial f}{\partial t} \right]_{z=\eta} + \left[ \frac{\partial (f u_1)}{\partial x_1} + \frac{\partial (f u_2)}{\partial x_2} \right]_{z=\eta} \\
&\qquad + \left[ \frac{\partial f}{\partial z} \right]_{z=\eta} \underbrace{\left( \frac{\partial \eta}{\partial t} + [u_1]_{z=\eta} \frac{\partial \eta}{\partial x_1} + [u_2]_{z=\eta} \frac{\partial \eta}{\partial x_2} \right)}_{Q_1} \\
&\qquad + [f]_{z=\eta} \underbrace{\left( \left[ \frac{\partial u_1}{\partial z} \right]_{z=\eta} \frac{\partial \eta}{\partial x_1} + \left[ \frac{\partial u_2}{\partial z} \right]_{z=\eta} \frac{\partial \eta}{\partial x_2} \right)}_{Q_2}.
\end{aligned} \tag{B8}$$

Now, invoking the free-surface kinematic condition, Eq. (49), we find that

$$Q_1 = [u_3]_{z=\eta} \qquad \text{and} \qquad Q_2 = \left[ \frac{\partial u_3}{\partial z} \right]_{z=\eta},$$

where the second relation is derived by differentiating the kinematic condition with respect to $z$. Using those relations in Eq. (B8) and rearranging terms, we obtain

$$\frac{\partial}{\partial t}\left( [f]_{z=\eta} \right) + \nabla_2 \cdot \left( [f \cdot (u_1, u_2)]_{z=\eta} \right) = \left[ \frac{\partial f}{\partial t} + \nabla \cdot (f\, \boldsymbol{u}) \right]_{z=\eta},$$

completing the proof.